\documentclass[letterpaper]{article}
\usepackage{aaai2027} % DO NOT CHANGE THIS: anonymous submission mode
\usepackage[hyphens]{url} % DO NOT CHANGE THIS
\usepackage{graphicx} % DO NOT CHANGE THIS
\usepackage{natbib} % DO NOT CHANGE THIS
\usepackage{caption} % DO NOT CHANGE THIS
\title{MASTraceBench: Diagnosing Collaboration Gains through Proposal Trajectories in LLM-Based Multi-Agent Systems}

\author{
Yapeng Li$^{1}$,
Songze Li$^{1}$,
Shuang Yu$^{1}$,
Jing Yu$^{1}$,
Zhixin Liu$^{1}$,
Liqiang Wen$^{2}$,
Tonghua Su$^{1}$\thanks{Corresponding author}
\\[0.5ex]
$^{1}$ Harbin Institute of Technology, Harbin, China \\
$^{2}$ Peking University, Beijing, China
\\[0.5ex]
\{liyapeng, lisongze, yujing, Zhixin\_Liu\}@stu.hit.edu.cn,
\{yushuang, thsu\}@hit.edu.cn,
wenlq@pku.edu.cn
}

\affiliations{
}

\usepackage{booktabs}
\usepackage{multirow}
\usepackage{xcolor}
\usepackage{refcount}
\usepackage{tabularx}
\usepackage{makecell}
\usepackage{amsmath}
\usepackage{amssymb}
\usepackage{pifont}
\usepackage{algorithm}
\usepackage{algorithmic}

\usepackage[table]{xcolor}
\definecolor{oursbg}{RGB}{242,242,242}

\begin{document}

\maketitle

\begin{abstract}
LLM-based multi-agent systems (MAS) have shown promise in complex problem solving. As MAS methods diversify, systematic evaluation becomes
increasingly challenging. 
However, existing benchmarks largely focus on final outcomes,
leaving unclear how collaboration gains arise, are preserved, or
are lost.
To address this limitation, we introduce \textbf{MASTraceBench}, a benchmark for diagnosing collaboration gains through proposal trajectories in MAS.
Across six cooperative and competitive tasks, MASTraceBench
tracks and grades proposal trajectories and provides a multi-layer
metric suite covering Task Score, Collaboration Gain,
proposal-trajectory indicators, and Token Cost.
Using MASTraceBench, we systematically compare representative
MAS methods not only by final performance, but also by how agent
proposals evolve and are aggregated into the final answer. This
analysis reveals a recurring pattern: final MAS answers rarely
surpass the strongest initial proposal; interaction often lifts
initially weaker proposals toward it, while strong initial proposals
are seldom further improved and may regress.
To reduce this risk, we propose \textbf{CLEARS}, which replaces
whole-proposal exchange with claim-level evaluation across agents to guide reliable synthesis. 
CLEARS more often preserves or improves upon the strongest
initial proposal and achieves the highest Collaboration Gain on
five of the six tasks.
\end{abstract}    
\begin{figure}[!t]
\centering
\includegraphics[width=\linewidth]{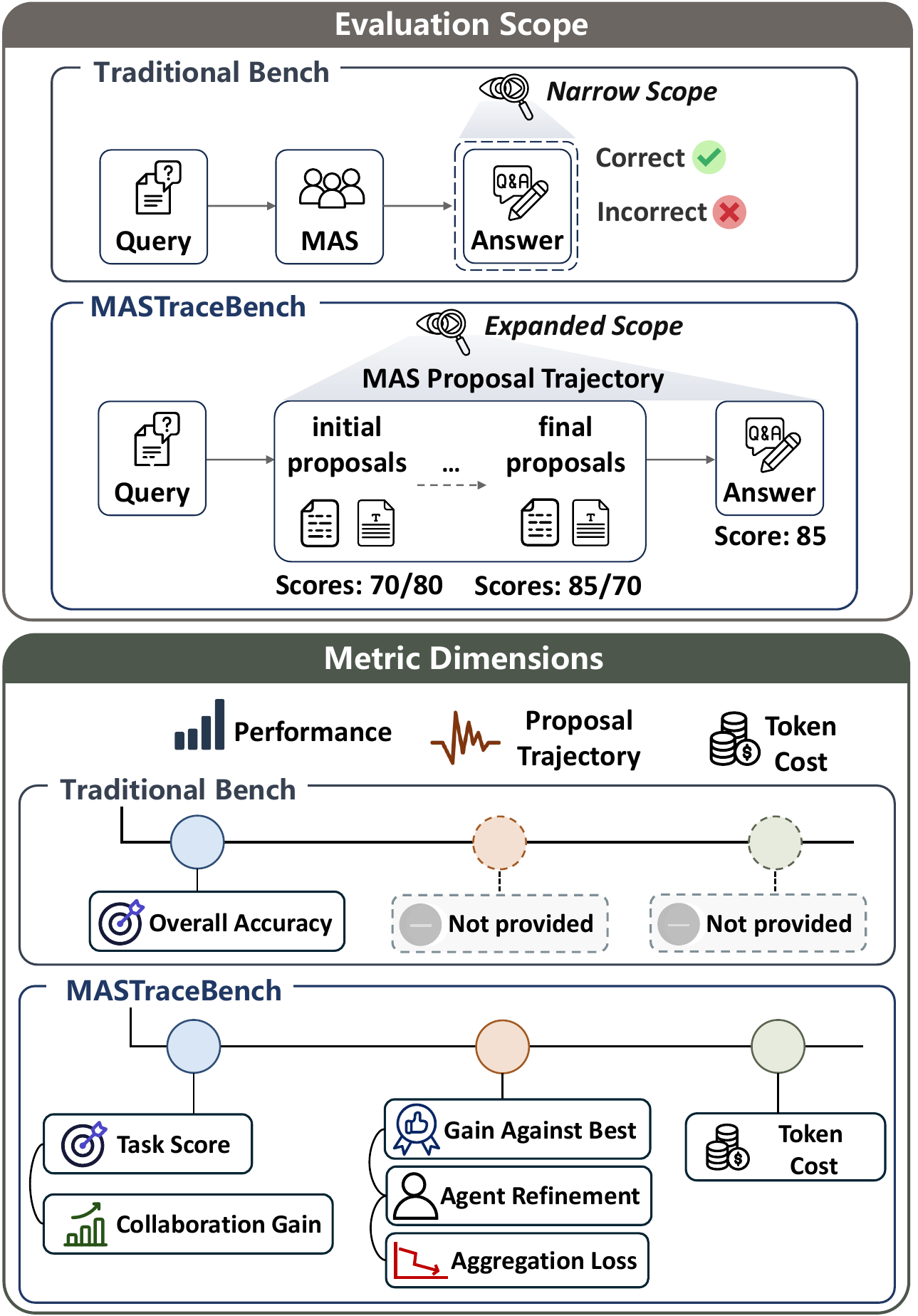}
\caption{
Comparison of evaluation scope and metric dimensions between traditional benchmarks and MASTraceBench.
}
\label{fig:MASTraceBench}
\end{figure}

\section{Introduction}
\label{sec:intro}

%----------LLM and MAS-------------------
Agents based on large language models (LLMs) have demonstrated strong capabilities in
understanding task instructions, reasoning and planning, and
taking actions toward task goals~\cite{wang2024survey,xi2025rise}.
Building on these capabilities, LLM-based multi-agent systems
(MAS) coordinate multiple agents with different roles, expertise,
or viewpoints to exchange information and jointly solve complex
tasks~\cite{wu2024autogen,hong2024metagpt,du2024improving}. 
As MAS methods become diverse in their agent
configurations, communication protocols, and decision mechanisms,
systematically evaluating them under comparable settings has
become essential for clarifying the practical value of MAS.

Reliable assessment of MAS requires evaluation approaches that
account for its collaborative nature. Despite growing efforts
in MAS evaluation, existing benchmarks offer limited diagnostic
granularity beyond final outcomes.
Most existing studies on LLM-based MAS~\cite{chen2024reconcile,fan2026cost} reuse general-purpose benchmarks for tasks such as mathematical reasoning~\citep{cobbe2021gsm8k} and code generation~\citep{chen2021evaluating}, retaining conventional protocols that judge only whether the final answer is correct, as illustrated in Fig.~\ref{fig:MASTraceBench}. 
This outcome-only evaluation obscures fine-grained improvements, such as transitions from incorrect to partially correct solutions, and makes it difficult to determine whether collaboration gains arise from the initial proposal pool, subsequent interaction, or final aggregation.
As fixed and widely reused test sets, these benchmarks also increasingly face performance saturation~\citep{akhtar2026ai} and potential training-data exposure~\citep{sainz2023nlp}, potentially biasing comparisons among
MAS methods.
Recent MAS benchmarks broaden task coverage to include coordination, competition, and sequential decision-making in multi-agent settings, with some further extending the evaluation metrics beyond final outcomes to assess intermediate processes~\citep{sun2025collab,zhu2025multiagentbench,lin2025gamebot}.
However, they are primarily designed to compare model backbones in multi-agent settings rather than to compare and diagnose MAS methods.
Their intermediate metrics do not directly track the trajectories of candidate solutions during interaction and aggregation, providing limited insight into how collaboration gains are produced, preserved, or lost.

%---------------------------our work--------------------------
To address this gap, we introduce \textbf{MASTraceBench}, a benchmark
for diagnosing MAS collaboration gains through proposal-trajectory
analysis, as illustrated in Fig.~\ref{fig:MASTraceBench}. We define a
\emph{proposal} as an evaluable candidate solution generated within
the MAS for the current query or state. Rather than evaluating only
the final MAS answer, MASTraceBench records the initial and final
proposals of Proposer Agents together with the final MAS answer,
thereby capturing proposal trajectories during interaction and
aggregation. It assigns graded scores to these outputs, making
fine-grained changes in proposal quality directly measurable.
Accordingly, MASTraceBench reports a broader metric suite beyond Task Score, covering Collaboration Gain, proposal-trajectory indicators, and Token Cost. It covers six cooperative and competitive tasks in which
proposals are observable and scoreable, enabling comparison across MAS methods.
% These metrics help diagnose how MAS gains are produced, preserved, or lost across proposal stages, while accounting for cost. 
% MASTraceBench covers six cooperative and competitive task scenarios, each designed to make proposal-level outputs observable and scoreable for method-level comparison.

% Using MASTraceBench, we evaluate representative MAS methods and reveal recurring patterns that are difficult to observe from final performance alone. Final MAS answers rarely surpass the strongest initial proposal; interaction more often lifts weaker proposals toward it, while strong proposals are seldom further improved and may regress. This suggests that current MAS gains largely rely on preserving and propagating existing strong solutions, whereas coarse whole-proposal interaction may also disturb their reliable components. 
% To mitigate strong-proposal regression while better integrating
% useful information across proposals, we introduce CLEARS, which
% replaces whole-proposal exchange with claim-level evaluation across agents to guide reliable synthesis.
% % Motivated by these findings, we introduce \textbf{CLEARS}, which replaces whole-proposal exchange with claim-level evaluation across agents to guide reliable synthesis.
% Experimental results show that CLEARS more often retains or surpasses the best initial proposal, while achieving the strongest performance on most benchmarks.

Using MASTraceBench, we evaluate representative MAS methods and
uncover a recurring pattern hidden by final performance: final MAS
answers rarely surpass the strongest initial proposal; interaction
more often lifts initially weaker proposals toward it, while
initially strong proposals are seldom further improved and may
regress. Rather than meeting the common expectation that
multi-agent interaction creates stronger proposals,
the evaluated methods often gain by preserving and propagating
existing strong proposals. This result highlights a key limitation
in how current interaction mechanisms use proposal diversity:
whole-proposal exchange transmits useful and unreliable components
together and may not reliably preserve strong proposals, as reflected
in the observed regressions. 
To reduce this risk, we introduce CLEARS, a method that replaces
whole-proposal exchange with claim-level evaluation across agents to guide reliable synthesis. 
CLEARS more often preserves or improves upon the strongest initial proposal and achieves the highest Collaboration Gain on five of the six benchmarks.

Our contributions are summarized as follows:
\begin{itemize}

\item We introduce \textbf{MASTraceBench}, which moves beyond
outcome-only MAS evaluation to enable fine-grained diagnosis of
collaboration gains. To the best of our knowledge, it is the first
benchmark to support proposal-trajectory diagnosis at the method
level.

\item Using MASTraceBench, we systematically evaluate
representative MAS methods across six tasks, revealing a
recurring pattern underlying collaboration gains: strong initial
proposals are rarely improved and may regress, whereas weaker
ones are often lifted toward them.

\item We introduce CLEARS, which enables selective
synthesis by preserving useful proposal components and avoiding
unreliable ones. CLEARS more reliably preserves or improves upon
the strongest initial proposal and achieves the highest
Collaboration Gain on five of the six benchmarks.

\end{itemize}

\section{Related Work}
\label{sec:relat}

\subsection{LLM-based Multi-Agent Systems}
% Recent advances in large language models (LLMs) have led to growing interest in MAS. 
LLM-based MAS have rapidly evolved, enabled by recent advances
in large language models.
Early work on self-consistency~\citep{wang2023selfconsistency} shows that aggregating multiple reasoning paths can improve performance. Reflection-based approaches~\citep{madaan2023self,shinn2023reflexion} further refine model outputs through iterative feedback.
Later work explores explicit interaction among multiple agents, including debate~\citep{du2024improving,liang2024encouraging}, peer review~\citep{xu2023towards}, and verification~\citep{lifshitz2025multi}. Another important line of work focuses on automatic or adaptive MAS construction, including AgentVerse~\citep{chen2024agentverse}, which dynamically adjusts multi-agent group composition, and MAS-GPT~\citep{ye2025masgpt}, which generates query-adaptive MASs from user queries. MASTraceBench evaluates these representative methods and provides diagnostics of their collaboration gains.

\subsection{MAS Benchmarks}
Benchmarks used in MAS studies can be broadly divided into two categories: general-purpose benchmarks reused for MAS evaluation and benchmarks specifically designed for multi-agent settings. General-purpose benchmarks such as MATH~\citep{hendrycks2021measuring},
HumanEval+~\citep{evalplus}, MMLU-Pro~\citep{wang2024mmlu}, and
GPQA~\citep{rein2024gpqa} cover mathematical reasoning, code
generation and multidomain reasoning, but typically assess only final-answer correctness. MAS-specific benchmarks broaden this scope: Collab-Overcooked~\citep{sun2025collab} and BattleAgentBench~\citep{wang2024battleagentbench} evaluate sequential collaboration and cooperative and competitive task execution in structured interactive environments, respectively. LLMArena~\citep{chen2024llmarena} spans diverse dynamic games to assess strategic planning, communication, opponent modeling, and team collaboration. MultiAgentBench~\citep{zhu2025multiagentbench} further measures task
progress and coordination quality through milestone, communication, and
planning scores, while GAMEBoT~\citep{lin2025gamebot} decomposes game
reasoning into predefined subproblems with rule-based intermediate
verification.
These benchmarks primarily compare model backbones in particular multi-agent settings, whereas MASTraceBench focuses on method-level diagnosis by tracking proposal trajectories during interaction and aggregation.

% MASLab further supports method-level comparison by unifying implementations of existing MAS methods~\citep{ye2025maslab}. 
% However, existing evaluations provide limited insight into how collaboration produces or loses improvements over candidate proposals.

\subsection{Sources and Reliability of MAS Gains}
Recent studies examine how MAS gains arise and under what conditions they remain reliable. Debate-or-Vote~\citep{choi2026debate} decomposes debate gains into majority voting and inter-agent discussion. A controlled study~\citep{wu2025can} examines the effects of initial reasoning strength, group diversity, and majority pressure, while MAST~\citep{pan2025multiagent} identifies recurrent MAS failure modes from execution traces. A scaling-oriented study~\citep{kim2025towards} further shows that MAS gains depend on model capability, task properties, and MAS architecture. MASTraceBench advances this research by quantitatively tracking proposal trajectories to reveal how collaboration gains are produced, preserved, or lost.

\section{MASTraceBench}
\label{sec:MASTraceBench}

\begin{figure*}[t]
    \centering
    \includegraphics[width=\linewidth]{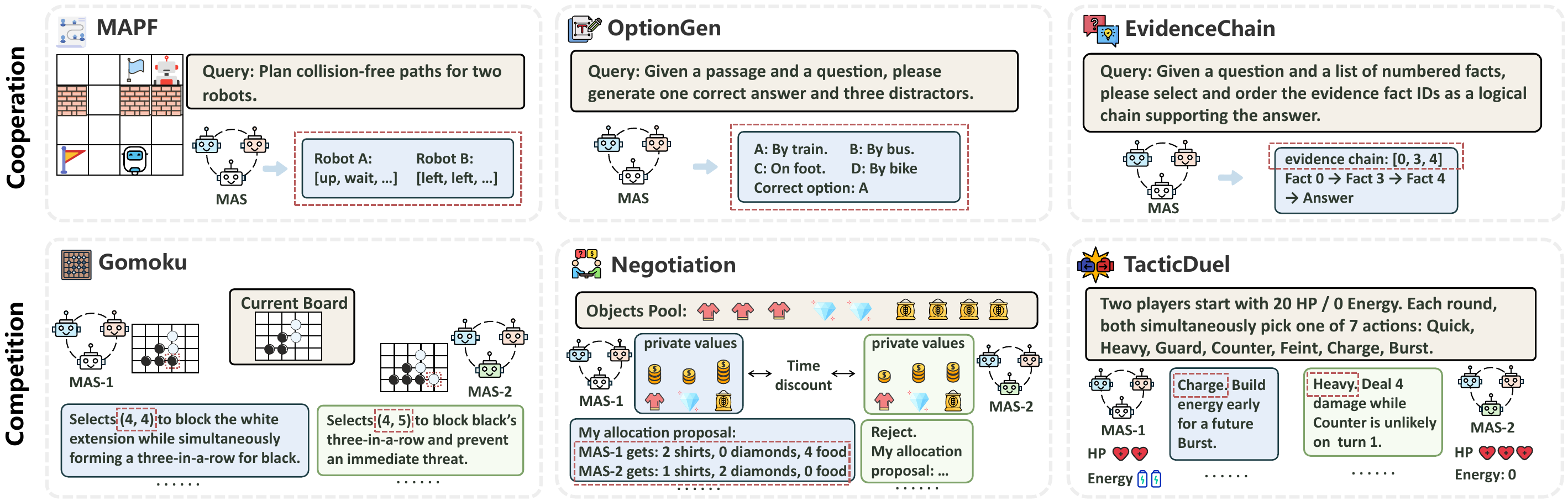}
    \caption{Benchmark design in MASTraceBench. Red dashed boxes indicate the proposal-level outputs evaluated by $S(\cdot)$.}
    \label{fig:benchmark}
\end{figure*}

\begin{table*}[t]
\centering
\small
\setlength{\tabcolsep}{5.0pt}
\renewcommand{\arraystretch}{1.08}
\begin{tabular*}{\textwidth}{@{\extracolsep{\fill}}lccp{7.2cm}}
\toprule
\textbf{Benchmark} & \textbf{Setting} & \textbf{Task Score (TS)} & \textbf{Graded Proposal Scoring Function $S(\cdot)$} \\
\midrule
MAPF & 157 samples & Mean arrival rate & Fraction of robots reaching their goals \\
OptionGen & 235 samples & Mean option-set score & LLM-judge score of the correct option and distractor set \\
EvidenceChain & 100 samples & Mean chain score & Set F1 and ordering agreement with ground truth \\
Gomoku & Dynamic & Win rate & Self-formation score plus defense-interception score \\
Negotiation & Dynamic & Mean episode payoff ratio & Payoff ratio of proposed allocation \\
TacticDuel & Dynamic & Win rate & Rule-based tactical value of the proposed action \\
\bottomrule
\end{tabular*}
\caption{Tasks and evaluation specifications in MASTraceBench. The first three tasks are static tasks that require one-shot solutions for independent samples, while tasks marked as Dynamic require MASs to repeatedly observe the current state and output actions until an episode ends. Additional implementation details are provided in the supplementary material.}
\label{tab:benchmark_design}
\end{table*}

We introduce MASTraceBench with
(1) a unified evaluation framework for consistently capturing
proposal trajectories across MAS methods and task scenarios;
(2) a multi-layer metric suite for diagnosing collaboration
gains beyond final outcomes; and
(3) a diverse benchmark suite spanning six cooperative and
competitive tasks with graded scoring along proposal trajectories.

\subsection{Unified Evaluation Framework}

To consistently capture and evaluate proposal trajectories across MAS methods and task scenarios, we organize the evaluation framework around agent modeling and environment modeling.

\paragraph{Agent Modeling.}
We distinguish agents by whether their outputs can be directly evaluated as proposals. 
Within an MAS, agents that produce candidate proposals for a given query or state are treated as \textit{Proposer Agents}. MASTraceBench records and scores their initial and final proposals together with the final MAS answer; these outputs constitute the MAS's \emph{proposal trajectory} for that query or state.
Agents that provide feedback, verification, critique, or judgment without producing such proposals are treated as \textit{Function Agents}, whose outputs serve as auxiliary interaction signals rather than independently evaluated proposals.

\paragraph{Environment Modeling.}
We model different benchmark tasks under a unified query--MAS--answer paradigm. For static tasks, the MAS receives a query and produces a final MAS answer, while MASTraceBench records the corresponding proposal trajectory. For dynamic tasks, the MAS repeatedly observes the current state, outputs an action, and receives the next state according to
\begin{equation}
(q, s_t) \rightarrow \mathrm{MAS} \rightarrow a_t,
\qquad
s_{t+1} = \mathcal{T}(s_t, a_t),
\end{equation}
where $q$ is the task query, $s_t$ is the environment state at step $t$, $a_t$ is the MAS answer at that step, and $\mathcal{T}$ is the transition function. MASTraceBench records a corresponding proposal trajectory for each action decision, and the complete sequence of actions determines the final task outcome.

\subsection{Multi-Layer Metrics}
\label{multi-layer metrics}
To address the limited diagnostic granularity of outcome-only evaluation, we define a multi-layer metric suite for diagnosing collaboration gains, with proposal-level diagnostics grounded in the recorded proposal trajectories.

%---------------------------------Layer-1---------------------------------

\paragraph{Layer-1: How much gain does the MAS achieve?}
We use \textbf{Task Score (TS)} to measure the task-level performance of a MAS:
\begin{equation}
TS = \mathbb{E}_{z \in \mathcal{Z}}
\left[ R_{\mathrm{MAS}}(z) \right],
\end{equation}
% where $\mathcal{Z}$ denotes the set of task-level instances. Each instance is an individual sample for static tasks or a complete episode for dynamic tasks, where the full action sequence determines the final task outcome. $R_{\mathrm{MAS}}(z)$ denotes the task score achieved by the MAS on instance $z$.

where $\mathcal{Z}$ is the set of task-level instances, i.e., samples for static tasks and episodes for dynamic tasks. $R_{\mathrm{MAS}}(z)$ is the final MAS answer score for static tasks and the episode score determined by the MAS action sequence for dynamic tasks.

To quantify the overall performance improvement achieved by a MAS over
the Single-Agent baseline, we define \textbf{Collaboration Gain
(CG)} as:
\begin{equation}
CG = TS - TS_{\mathrm{Single}},
\end{equation}
where $TS_{\mathrm{Single}}$ denotes the Task Score achieved by the Single-Agent baseline under the same benchmark and evaluation protocol.

%---------------------------------Layer-2---------------------------------

\paragraph{Layer-2: How does the MAS produce, preserve, or lose gains?}
%-----------example-----------------------
A final MAS answer score alone can hide very different proposal trajectories.
Consider a MAPF instance where scores are arrival rates and, in all cases, the final MAS answer scores $66.7\%$.
If all initial proposals score $33.3\%$, this answer indicates newly created value beyond the initial proposal pool.
If the initial proposals score $100.0\%$, $33.3\%$, and $33.3\%$, the same answer instead fails to preserve the strongest initial proposal, even though some individual proposals may have been refined from $33.3\%$ to $66.7\%$ during collaboration.
If a final Proposer proposal still scores $100.0\%$ but the aggregated MAS answer scores $66.7\%$, the loss occurs during aggregation.
These cases motivate GAB, AR, and AL, which respectively compare the final MAS answer with the best initial proposal, track the initial-to-final evolution of individual Proposer proposals, and measure aggregation loss.

We formalize these proposal trajectory diagnostics over a set of evaluation units $\mathcal{D}$, where each $x \in \mathcal{D}$ corresponds to a sample in a static task or a state-conditioned action decision in a dynamic task. For each $x$, let $S_{\mathrm{MAS}}(x)$ denote the score of the final MAS answer. Let $\mathcal{P}_m$ denote the set of Proposer Agents in MAS method $m$, and let $S_i^{\mathrm{init}}(x)$ and $S_i^{\mathrm{final}}(x)$ denote the scores of agent $i$'s initial and final proposals, respectively. 
For subsequent diagnostics, we define
$b_x = \max_{i \in \mathcal{P}_m} S_i^{\mathrm{init}}(x)$
as the best initial proposal score and let $S_x^{\star}$ denote the maximum attainable score for $x$.

%----------------GAB---------------------
\textbf{Gain Against Best (GAB).} 
% We use GAB to characterize how the final MAS answer changes relative to the best initial proposal. 
For each $x \in \mathcal{D}$, we define GAB label as:

\[
\mathrm{GAB}(x)=
\left\{
\begin{array}{@{}l@{\;}l@{}}
\textsc{Improve}, &
S_{\mathrm{MAS}}(x)>b_x,\\
\textsc{Regress}, &
S_{\mathrm{MAS}}(x)<b_x,\\
\textsc{OptKeep}, &
S_{\mathrm{MAS}}(x)=b_x=S_x^\star,\\
\textsc{SubKeep}, &
S_{\mathrm{MAS}}(x)=b_x<S_x^\star.
\end{array}
\right.
\]

% These labels indicate whether the MAS surpasses the best initial proposal, falls below it, preserves an optimal initial best, or retains a suboptimal initial best without further improvement.
The overall GAB is reported as the percentage distribution of these labels over $\mathcal{D}$.

%----------AR--------------------------
\textbf{Agent Refinement (AR).} 
% GAB captures changes relative to the best initial proposal, but it does not reveal how individual Proposer Agents contribute to these changes. 
% We therefore define AR to track proposal evolution from 
We define AR to track proposal evolution from 
$S_i^{\mathrm{init}}(x)$ to $S_i^{\mathrm{final}}(x)$.
To examine how proposal refinement varies with initial quality, we partition agents \emph{on each instance $x$} into Strong Agents and Weak Agents according to their initial proposal scores: Strong Agents are those in $\mathcal{B}_x = \{i \in \mathcal{P}_m \mid S^{\text{init}}_i(x) = b_x\}$, while Weak Agents are those in $\mathcal{W}_x = \mathcal{P}_m \setminus \mathcal{B}_x$.

For agents in $\mathcal{B}_x$, their initial score is $b_x$. 
We define their refinement label $\text{AR}_{\text{strong}}(x, i)$ using the same four conditions as $\text{GAB}(x)$, simply by replacing $S_{\text{MAS}}(x)$ with $S^{\text{final}}_i(x)$.

For agents in $\mathcal{W}_x$, we measure their refinement relative to both their own initial score and the strong baseline $b_x$. 
Their refinement label is defined as:

\[
\mathrm{AR}_{\mathrm{weak}}(x,i)=
\left\{
\begin{array}{@{}l@{\;}l@{}}
\textsc{Surpass}, &
S_i^{\mathrm{final}}(x)>b_x,\\
\textsc{Reach}, &
S_i^{\mathrm{final}}(x)=b_x,\\
\textsc{Partial}, &
S_i^{\mathrm{init}}(x)<S_i^{\mathrm{final}}(x)<b_x,\\
\textsc{Keep}, &
S_i^{\mathrm{final}}(x)=S_i^{\mathrm{init}}(x),\\
\textsc{Degrade}, &
S_i^{\mathrm{final}}(x)<S_i^{\mathrm{init}}(x).
\end{array}
\right.
\]

% For weak agents, these labels respectively denote surpassing the best initial proposal, reaching it, improving without reaching it, remaining unchanged, and falling below the agent's own initial score.
For each AR group, we report label percentages within that group.

%-----------AL--------------------------
\textbf{Aggregation Loss (AL).} 
Even after proposal refinement, gains may still be lost when the final MAS answer is aggregated.
We therefore define AL as the gap between the final MAS answer and the best final proposal:

\begin{equation}
\begin{gathered}
AL =
\mathbb{E}_{x \in \mathcal{D}}
\left[
\max_{i \in \mathcal{P}_m}
S_i^{\mathrm{final}}(x)
-
S_{\mathrm{MAS}}(x)
\right],
\\
\end{gathered}
\end{equation}

%---------------------------------Layer-3---------------------------------

\paragraph{Layer-3: How costly are these gains?}
Since MAS gains may rely on additional agent calls, interaction rounds, or verification steps, we use \textbf{Total Token Cost (TC)} to measure the average token consumption required by a MAS:

\begin{equation}
TC = \mathbb{E}_{x \in \mathcal{D}} \left[ C_{\mathrm{MAS}}(x) \right],
\end{equation}
where $C_{\mathrm{MAS}}(x)$ denotes the total token cost incurred by all agent calls in the MAS from receiving the input context to producing the corresponding MAS answer.

\subsection{Benchmark Design}

General-purpose benchmark tasks often require an MAS to
select an option or output a final numerical answer, so the
resulting outputs are ultimately judged only as correct or
incorrect. An improved but still incorrect proposal therefore remains
indistinguishable from its predecessor, while quality
differences among correct proposals are also hidden. To capture
the proposal-quality differences required for proposal-trajectory
analysis, we construct six evaluation tasks.

As shown in Fig.~\ref{fig:benchmark}, MASTraceBench
includes multi-agent path finding, option generation,
evidence-chain reasoning, board-game move selection,
private-value negotiation, and turn-based tactical
decision-making. These tasks span cooperative and competitive,
static and dynamic settings: a single MAS solves cooperative
tasks, whereas multiple MASs interact head-to-head in
competitive tasks. Each task assigns graded scores to
intermediate proposals and final MAS answers. For example, in
MAPF, guiding two rather than one of three robots to their goals
raises a proposal's score from 33.3\% to 66.7\%. Such scoring
makes fine-grained changes in proposal quality observable and
supports proposal-trajectory analysis across diverse forms of
multi-agent interaction. Table~\ref{tab:benchmark_design}
summarizes each task's setting, Task Score, and proposal-level
scoring function $S(\cdot)$.

\subsection{CLEARS: Claim-Level Evaluation Across Agents for Reliable Synthesis}

\begin{figure}[t]
    \centering
    \includegraphics[width=\columnwidth,keepaspectratio]{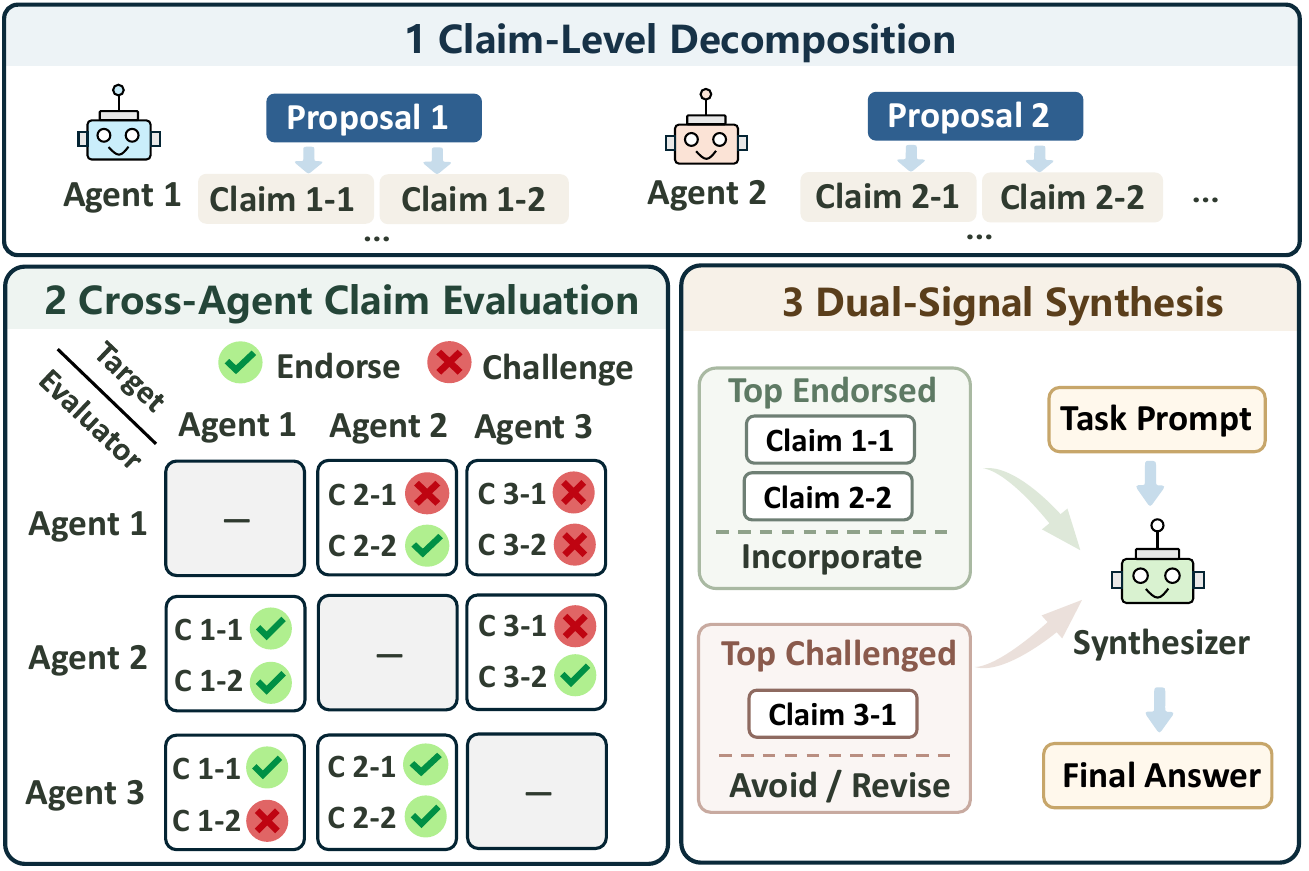}
    \caption{
        Overview of CLEARS.
    }
    \label{fig:CLEARS}
\end{figure}

% Motivated by the strong-proposal regression in our analysis, we introduce \textbf{Claim-Challenge
% Synthesis (CCS)}, a diagnosis-guided MAS method that shifts
% interaction from whole proposals to structured claims, as
% illustrated in Fig.~\ref{fig:ccs}.
% After generating its proposal, each Proposer Agent is asked to state a set of supporting claims, each capturing a key reasoning or decision unit in its proposal.
% For each claim, other agents provide \emph{endorsement} signals for reliable or useful claims and \emph{challenge} signals for unreliable or contested ones.
% A Synthesizer Agent then uses endorsed and challenged claims as positive and negative signals to produce the final MAS answer. This claim-level design aims to preserve useful components while reducing interference from coarse proposal-level revision or aggregation.

To reduce the risk caused by propagating useful and unreliable
proposal components together, we introduce \textbf{CLEARS}, a diagnosis-guided MAS method that shifts interaction and synthesis from whole proposals to claim-level signals.

For an input $x$, let
$\mathcal{Y}_x=\{y_i\mid i\in\mathcal{P}_m\}$
denote the proposals generated by the Proposer Agents. For analysis, we view each proposal as consisting of semantically meaningful solution components, such as factual assertions, reasoning steps, or local decisions:
\begin{equation}
\phi(y_i)=(\mathcal{U}_i,\mathcal{E}_i),
\end{equation}
where $\mathcal{U}_i$ and $\mathcal{E}_i$ denote latent components that contribute positively and negatively to the overall quality of proposal $y_i$, respectively.

In whole-proposal interaction, Agent $i$ revises its proposal by conditioning on the other proposals in their entirety:
\begin{equation}
\tilde{y}_i
\sim
p_\theta\!\left(
\cdot\mid x,y_i,\{y_j\}_{j\neq i}
\right).
\end{equation}

Providing each $y_j$ intact leaves $\mathcal{U}_j$ and
$\mathcal{E}_j$ undifferentiated during revision, while subsequent
majority voting or LLM-based selection may overlook
$\mathcal{U}_j$ when the corresponding proposal is not selected.

As shown in Fig.~\ref{fig:CLEARS}, CLEARS
translates this latent distinction into a three-stage procedure.
Each Proposer Agent first decomposes its proposal into claims,
each explicitly representing a solution component. Cross-agent evaluation aggregates different agent perspectives
to identify the top-endorsed claims $\mathcal{C}_x^{+}$ and
top-challenged claims $\mathcal{C}_x^{-}$, which respectively
approximate the useful components
$\{\mathcal{U}_i\}_{i\in\mathcal{P}_m}$ and unreliable components
$\{\mathcal{E}_i\}_{i\in\mathcal{P}_m}$ across proposals. The final MAS answer is then
generated as
\begin{equation}
\hat{y}_{\mathrm{CLEARS}}
\sim
p_\theta\!\left(
\cdot\mid x,\mathcal{C}_x^{+},\mathcal{C}_x^{-}
\right).
\end{equation}
During synthesis, $\mathcal{C}_x^{+}$ provides content to
preserve or combine, whereas $\mathcal{C}_x^{-}$ identifies
content to avoid, enabling CLEARS to construct a new final
MAS answer from information distributed across proposals.
\section{Experiments}
\label{sec:experiments}

\subsection{Setup}
\label{setup}

\paragraph{Evaluated Methods.}
We evaluate representative MAS methods from four categories (Table~\ref{tab:main_ts_cg}): selection-based, debate-based, revision-based, and adaptive MAS construction. The first three capture common collaboration primitives, while the last automatically configures agent roles and workflows. All methods use Qwen3-32B~\citep{yang2025qwen3} as the shared backbone. 
For methods with configurable multi-Proposer settings, where
multiple Proposer Agents generate candidate proposals, we use
three Proposers; Int.~Debate and MAPR use one interaction round.

\paragraph{Evaluation Settings.}
For static tasks, we report the main results as averages over three runs per method. 
For competitive tasks, each method plays a fixed Single-Agent
baseline, avoiding opponent-set-dependent tournaments. We
conduct 50 Gomoku and TacticDuel games and evaluate 100
Negotiation configurations, each defined by an item pool and
two private-value profiles. To control order effects, we alternate the first mover in Gomoku and run each Negotiation configuration twice, reversing which side makes the first offer.
\subsection{Results and Analysis}
\label{sec:results}

We present the main results on MASTraceBench and then provide a detailed analysis of representative MAS methods.

%-------main results-----------
\begin{table*}[t]
\centering
\small
\setlength{\tabcolsep}{4.2pt}
\renewcommand{\arraystretch}{1.12}
\begin{tabular}{l*{12}{c}}
\toprule
\multirow{2}{*}{\textbf{Method}}
& \multicolumn{2}{c}{\textbf{MAPF}}
& \multicolumn{2}{c}{\textbf{OptionGen}}
& \multicolumn{2}{c}{\textbf{EvidenceChain}}
& \multicolumn{2}{c}{\textbf{Gomoku}}
& \multicolumn{2}{c}{\textbf{Negotiation}}
& \multicolumn{2}{c}{\textbf{TacticDuel}} \\
\cmidrule(lr){2-3}
\cmidrule(lr){4-5}
\cmidrule(lr){6-7}
\cmidrule(lr){8-9}
\cmidrule(lr){10-11}
\cmidrule(lr){12-13}
& \textbf{TS} & \textbf{CG}
& \textbf{TS} & \textbf{CG}
& \textbf{TS} & \textbf{CG}
& \textbf{TS} & \textbf{CG}
& \textbf{TS} & \textbf{CG}
& \textbf{TS} & \textbf{CG} \\
\midrule
Single-Agent
& 69.66 & --
& 80.56 & --
& 74.02 & --
& 50.00 & --
& 41.00 & --
& 50.00 & -- \\
\midrule
Consistency~\citep{wang2023selfconsistency}
& 72.91 & 3.25
& 81.83 & 1.27
& 75.19 & 1.17
& 50.00 & 0.00
& 42.33 & 1.33
& 55.00 & 5.00 \\
MAV~\citep{lifshitz2025multi}
& 77.13 & 7.47
& 82.19 & 1.63
& 74.40 & 0.38
& 46.00 & -4.00
& 43.21 & 2.21
& 54.00 & 4.00 \\
\midrule
Int.~Debate~\citep{du2024improving}
& \underline{77.31} & \underline{7.65}
& 83.70 & 3.14
& \underline{77.94} & \underline{3.92}
& \textbf{80.00} & \textbf{30.00}
& 44.08 & 3.08
& \underline{64.00} & \underline{14.00} \\
Adv.~Debate~\citep{liang2024encouraging}
& 67.50 & -2.16
& 81.32 & 0.76
& 65.83 & -8.19
& 48.00 & -2.00
& 43.28 & 2.28
& 21.00 & -29.00 \\
\midrule
Reflection~\citep{renze2024self}
& 75.24 & 5.58
& 81.22 & 0.66
& 75.75 & 1.73
& 58.00 & 8.00
& 42.51 & 1.51
& 36.00 & -14.00 \\
MAPR~\citep{xu2023towards}
& 76.33 & 6.67
& \underline{84.03} & \underline{3.47}
& 77.81 & 3.79
& 72.00 & 22.00
& \underline{44.75} & \underline{3.75}
& 35.00 & -15.00 \\
\midrule
AgentVerse~\citep{chen2024agentverse}
& 76.04 & 6.38
& 81.62 & 1.06
& 76.46 & 2.44
& 64.00 & 14.00
& 41.78 & 0.78
& 52.00 & 2.00 \\
MAS-GPT~\citep{ye2025masgpt}
& 71.92 & 2.26
& 81.56 & 1.00
& 77.33 & 3.31
& 68.00 & 18.00
& 42.19 & 1.19
& 60.00 & 10.00 \\
\midrule
\textbf{CLEARS (ours)}
& \textbf{78.72} & \textbf{9.06}
& \textbf{84.55} & \textbf{3.99}
& \textbf{80.44} & \textbf{6.42}
& \underline{76.00} & \underline{26.00}
& \textbf{45.24} & \textbf{4.24}
& \textbf{66.00} & \textbf{16.00} \\
CLEARS w/o CCE
& 74.52 & 4.86
& 82.65 & 2.09
& 79.58 & 5.56
& 72.00 & 22.00
& 42.74 & 1.74
& 53.00 & 3.00 \\
\bottomrule
\end{tabular}
\caption{
Performance of MAS methods across benchmarks. 
TS and CG denote Task Score (0--100 scale) and Collaboration Gain over
the Single-Agent baseline; Int.~Debate and Adv.~Debate refer to
interactive and adversarial debate.
}
\label{tab:main_ts_cg}
\end{table*}

\begin{figure*}[t]
    \centering
    \includegraphics[width=\linewidth]{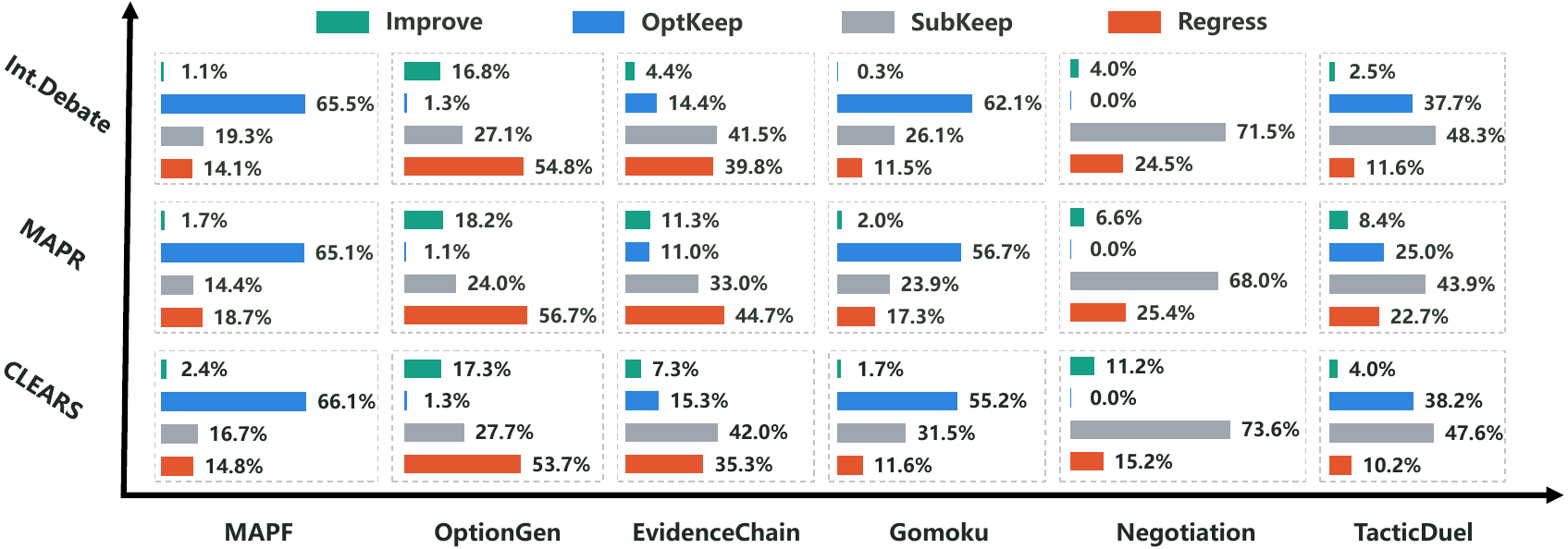}
    \caption{
    GAB analysis. Each bar reports the percentage distribution across GAB labels.
    % GAB is reported for methods with multiple Proposer Agents whose final MAS answer is not restricted to selecting from the initial proposal pool.
    % These categories indicate whether the final MAS answer surpasses the initial best proposal, falls below it, preserves it when already optimal, or preserves it when still improvable. GAB is reported for protocols with multiple Proposer Agents whose final MAS answer is not restricted to selecting from the initial proposal pool.
    }
    \label{fig:gab_ratio}
\end{figure*}

%-------------------------------------------------------------------------

%-------------------------------------------------------------------------

% \setlength{\floatsep}{10pt plus 1pt minus 1pt}
% \setlength{\textfloatsep}{12pt plus 1pt minus 2pt}

%--------------------------------------------------

\paragraph{Observation 1: Multi-proposer interaction brings larger gains in MAS.}
As shown in Table~\ref{tab:main_ts_cg}, MAS methods exhibit clear differences in gains over the Single-Agent baseline.
Interactive protocols with multiple Proposer Agents, such as Int.~Debate and MAPR, generally achieve stronger TS and CG than passive aggregation, external verification, or single-proposer refinement methods.
For example, on Gomoku, Int.~Debate reaches 80.00, outperforming Consistency (50.00), MAV (46.00), and Reflection (58.00), while MAPR also achieves a strong score of 72.00.
Adaptive frameworks such as AgentVerse and MAS-GPT perform reasonably on several benchmarks, but they do not consistently match the strongest interactive protocols, suggesting that workflow adaptation alone does not guarantee the most suitable collaboration mechanism.
Notably, Adv.~Debate's poor performance may stem from noisy
counter-proposals induced by its forced pro--con roles.
% Revision-based methods also perform poorly on TacticDuel, suggesting that extra revision may disturb initially strong action choices when the preferred action is relatively clear.

\paragraph{Observation 2: Interaction among multi-proposers produces limited new gains in MAS.}
Fig.~\ref{fig:gab_ratio} reports GAB for this class of methods, for which cross-proposer debate or revision would intuitively be expected to create gains beyond the strongest initial proposal.
However, the \emph{Improve} ratio remains limited on most benchmarks, showing that new gains are rare.
Most cases instead fall into \emph{OptKeep} or \emph{SubKeep}, suggesting that observed gains mainly come from carrying forward an already available strong proposal rather than forming a stronger answer.
The non-negligible \emph{Regress} ratios, especially on OptionGen and EvidenceChain, further show that even this preservation is not guaranteed.

\begin{figure}[t]
    \centering
    \includegraphics[width=\linewidth]{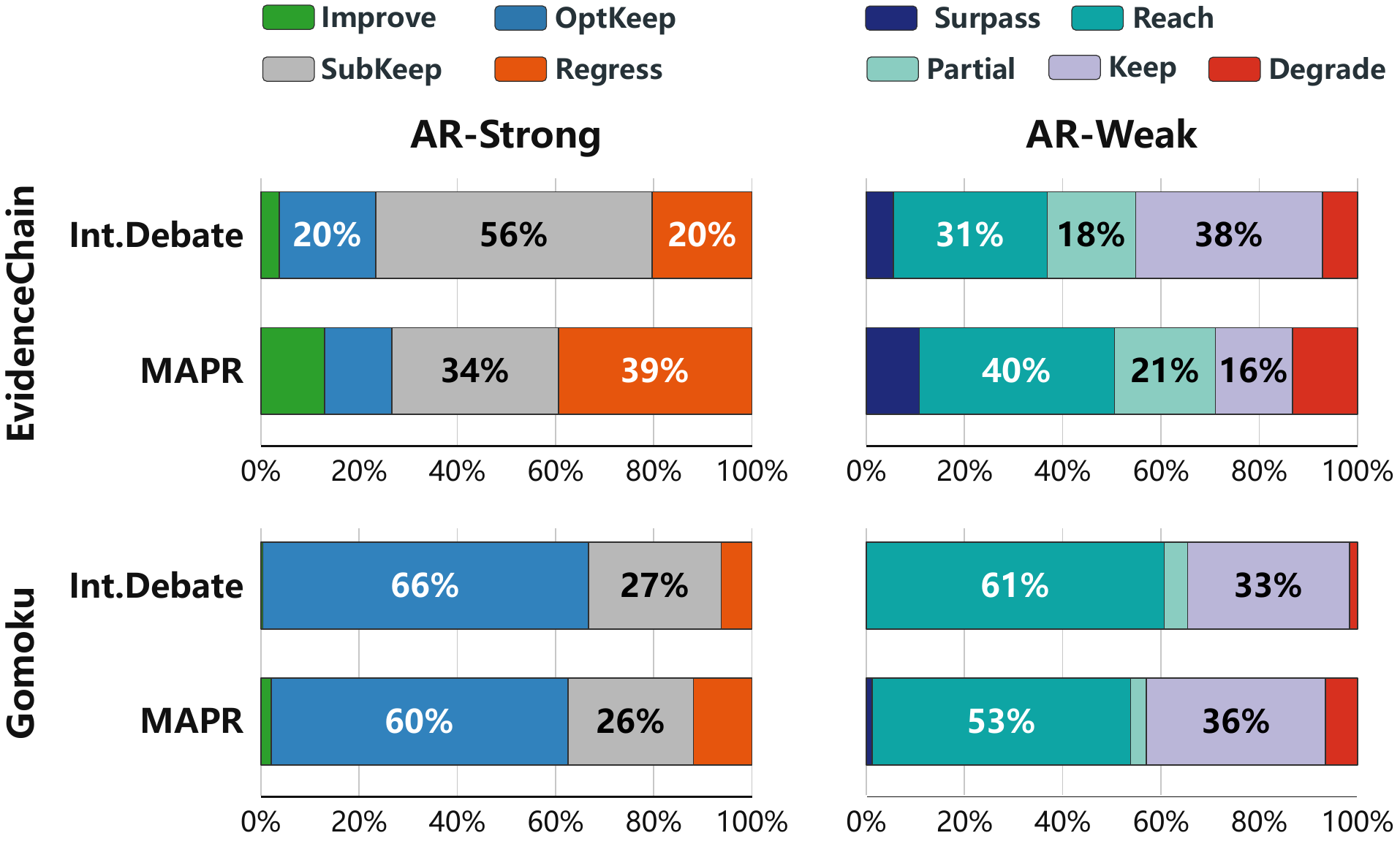}
    \caption{
    AR analysis on EvidenceChain and Gomoku. 
    For each sample, Proposer Agents are grouped as AR-Strong if their initial proposal achieves the best initial score.
    }
    \label{fig:ar}
\end{figure}

\paragraph{Observation 3: Weak Agents are lifted while Strong Agents are not reliably preserved.}
% Fig.~\ref{fig:ar} reports AR for Int.Debate and MAPR as two representative proposal-update modes: debate-based refinement and critique--revision. 
Fig.~\ref{fig:ar} further examines Observation~2 at the agent level.
Weak Agents often fall into \emph{Reach} or \emph{Partial}, moving toward
the strongest initial proposal but rarely beyond it.
For Strong Agents, \emph{Improve} remains limited and \emph{Regress} is non-negligible, indicating that initially strong proposals are hard to further improve and are not always preserved during interaction. 
This points to a limitation of whole-proposal interaction: agents exchange full proposals or full-proposal critiques, and whole-proposal revision may carry errors along with useful information.
For example, noisy signals from weaker proposals may be absorbed during revision and interfere with otherwise strong proposals. 
Such strong-proposal degradation is an important source of the GAB Regress observed in Observation~2.

% The comparison between Int.Debate and MAPR further illustrates the effect of update mode. 
% Int.Debate tends to promote convergence by exchanging and comparing existing proposals, whereas MAPR introduces stronger critique--revision signals that create more opportunities for improvement but also higher risk of regression.

\begin{figure}[t]
    \centering
    \includegraphics[width=\linewidth]{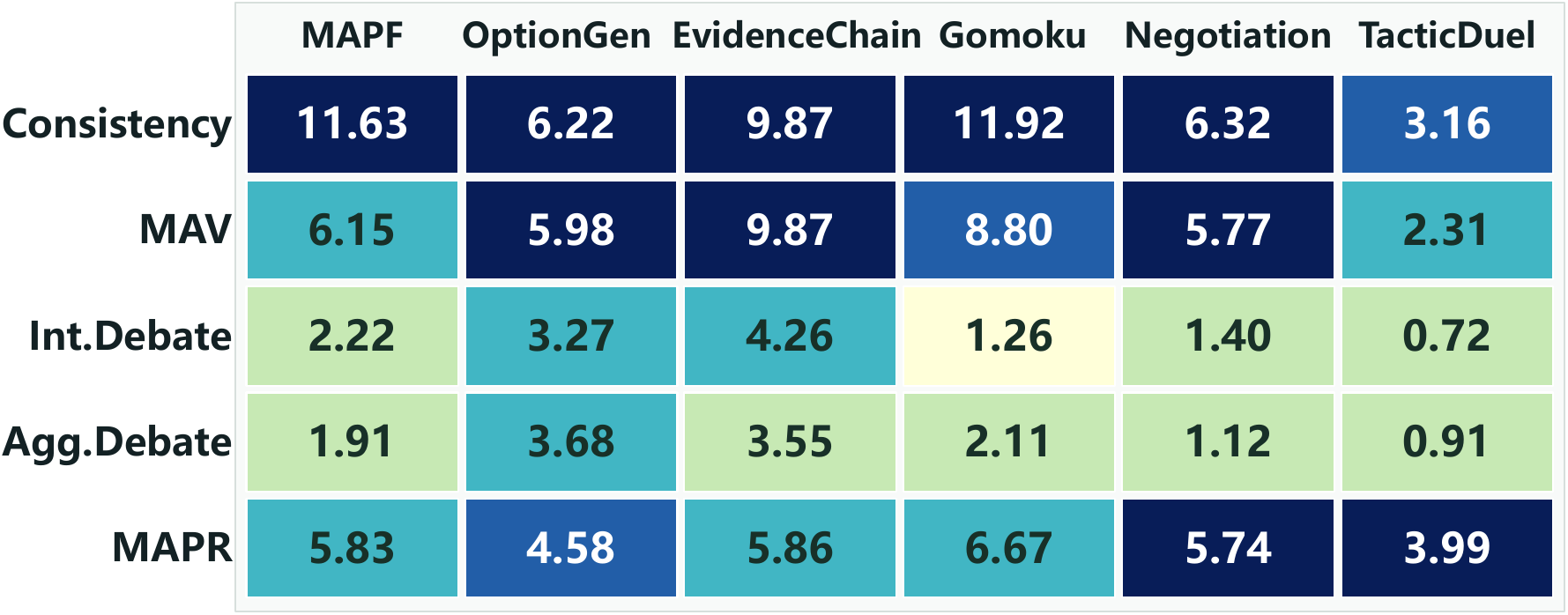}
    \caption{AL analysis. Lower AL indicates stronger aggregation capability of the MAS.}
    \label{fig:al}
\end{figure}

\paragraph{Observation 4: Aggregation loss depends on proposal convergence, not only the aggregation rule.}
Fig.~\ref{fig:al} shows that the final MAS answer can fall below the strongest final proposal among Proposer Agents.
Selection-based methods suffer large AL when strong proposals remain isolated. For example, on MAPF, Consistency has only 3.5\% final-proposal agreement and an AL of 11.63, while Int.~Debate raises this agreement to 58.2\% and reduces AL to 2.22.
Thus, effective aggregation requires proposal convergence before final decision, not merely a stronger aggregation rule.
Agg.~Debate further supports this view: replacing Int.~Debate's
majority-vote aggregation with LLM-based aggregation yields
comparable rather than uniformly better AL.

\paragraph{Observation 5: Token cost reveals the efficiency of MAS gains.}
Extra tokens support sampling, critique, verification, and revision,
but are valuable only when they yield collaboration gains. Taken
together, Tables~\ref{tab:main_ts_cg} and~\ref{tab:tc} show that,
compared with Consistency, Int.~Debate introduces debate among agents
and converts the additional token cost into stronger CG. In contrast,
the substantially costlier MAV and MAPR do not consistently yield
further gains. 
CLEARS incurs relatively higher TC, much of which is
attributable to cross-agent claim evaluation. The subsequent
component ablation isolates this effect through CLEARS w/o CCE:
adding CCE increases TC but consistently improves CG, indicating
that the added token budget is effectively converted into Collaboration Gain.

\begin{table}[t]
\centering
\small
\setlength{\tabcolsep}{7pt}
\begin{tabular}{lcc}
\toprule
\textbf{Method} & \textbf{MAPF} & \textbf{Gomoku} \\
\midrule
Single-Agent & $1.00\times$ & $1.00\times$ \\
Reflection   & $2.40\times$ & $2.38\times$ \\
Consistency  & $3.04\times$ & $3.04\times$ \\
Int.~Debate   & $5.14\times$ & $4.64\times$ \\
AgentVerse   & $5.75\times$ & $6.35\times$ \\
Adv.~Debate   & $5.79\times$ & $5.92\times$ \\
MAS-GPT      & $6.16\times$ & $6.23\times$ \\
MAV          & $8.62\times$ & $9.50\times$ \\
MAPR         & $10.50\times$ & $10.63\times$ \\
\midrule
CLEARS (ours)
             & $8.52\times$ & $8.38\times$ \\
CLEARS w/o CCE
             & $5.71\times$ & $5.53\times$ \\
\bottomrule
\end{tabular}
\caption{Token cost multipliers over the Single-Agent baseline
on MAPF and Gomoku.}
\label{tab:tc}
\end{table}

%---------------abl table--------------------
\begin{table}[t]
\centering

{\small
\setlength{\tabcolsep}{1.5pt}
\renewcommand{\arraystretch}{1.05}

\begin{tabular*}{\columnwidth}{
@{\extracolsep{\fill}}
lcccc@{\hspace{2pt}}cccc@{}
}
\toprule
\multirow{2}{*}{\textbf{Setting}}
& \multicolumn{4}{c}{\textbf{Int.~Debate}}
& \multicolumn{4}{c}{\textbf{MAPR}} \\
\cmidrule(lr){2-5}
\cmidrule(lr){6-9}
& \textbf{GI$\uparrow$}
& \textbf{GR$\downarrow$}
& \textbf{WI$\uparrow$}
& \textbf{SR$\downarrow$}
& \textbf{GI$\uparrow$}
& \textbf{GR$\downarrow$}
& \textbf{WI$\uparrow$}
& \textbf{SR$\downarrow$} \\
\midrule
Main (3P+1R)
& 4.4  & 39.8 & 54.9 & 20.3
& 11.3  & 44.7 & 71.1 & 39.4 \\
5P+1R
& 3.0  & 49.0 & 56.2 & 20.1
& 9.0  & 50.0 & 63.1 & 39.9 \\
7P+1R
& 4.0  & 53.0 & 57.4 & 30.5
& 3.0  & 50.0 & 61.1 & 37.5 \\
3P+2R
& 11.0 & 48.0 & 63.4 & 39.7
& 16.0 & 38.0 & 72.8 & 33.8 \\
3P+3R
& 7.0  & 41.0 & 70.6 & 32.9
& 6.0  & 42.0 & 67.9 & 34.8 \\
5P+2R
& 4.0  & 45.0 & 66.2 & 31.1
& 9.0  & 45.0 & 61.3 & 33.6 \\
Scale-Het.
& 7.0  & 40.0 & 55.9 & 23.8
& 7.0  & 46.0 & 68.1 & 48.2 \\
Family-Het.
& 15.0 & 61.0 & 59.8 & 50.5
& 11.0 & 72.0 & 53.8 & 63.1 \\
\bottomrule
\end{tabular*}
}

\caption{Configuration ablation on EvidenceChain. P and R
denote the numbers of Proposers and interaction rounds,
respectively; GI/GR denote GAB Improve/Regress; WI/SR denote
Surpass+Reach+Partial and Regress for initially weak and strong
agents, respectively. Under 3P+1R, Scale-Het. and Family-Het.
denote scale- and family-heterogeneous teams composed of
Qwen3-\{32B,14B,8B\} and \{Qwen3-8B, Llama-3.1-8B-Instruct,
Mistral-8B-Instruct\}, respectively.}

\label{tab:ablation_diag}
\end{table}

\paragraph{CLEARS Results.}
Table~\ref{tab:main_ts_cg} shows that CLEARS achieves the highest CG on five of the six benchmarks. On EvidenceChain, CLEARS reaches a CG of 6.42, exceeding the strongest baseline, Int.~Debate (3.92), by 2.50 points. The GAB results show a clearer advantage on Negotiation: CLEARS increases the \emph{Improve} ratio to 11.2\%, compared with 4.0\% for Int.~Debate and 6.6\% for MAPR, while reducing \emph{Regress} to 15.2\% from 24.5\% and 25.4\%, respectively (Fig.~\ref{fig:gab_ratio}). These results indicate that CLEARS improves collaboration gains
while mitigating proposal regression. 
Nevertheless, the \emph{Improve} ratio remains limited, and surpassing the strongest initial proposal remains an open challenge.

\section{Ablation Study}
\label{sec:ablation}

% We conduct ablation studies to evaluate the cross-agent
% endorsement--challenge component of CLEARS and the robustness
% of Observations~2--3 under varying interaction scales and
% model compositions.

% \paragraph{CLEARS component ablation.}
% CLEARS w/o CCE bypasses cross-agent claim evaluation and
% synthesizes directly from all decomposed claims. Its consistently
% lower CG across all six benchmarks
% (Table~\ref{tab:main_ts_cg}) confirms the value of CCE.

\paragraph{CLEARS component ablation.}
We compare CLEARS with CLEARS w/o CCE, which bypasses
cross-agent claim evaluation and directly synthesizes all
decomposed claims without filtering. Its consistently lower CG across all six benchmarks (Table~\ref{tab:main_ts_cg}) confirms the
contribution of CCE.

\paragraph{Interaction scale and model heterogeneity.}
We vary the number of Proposers, interaction rounds and model composition for Int.~Debate and MAPR
on EvidenceChain. As shown in Table~\ref{tab:ablation_diag},
GAB Improve remains limited across all configurations,
whereas GAB Regress and strong-proposal regression remain
substantial despite frequent improvements to initially weak
proposals. These results suggest that Observations~2--3 are robust across
configurations.

\section{Conclusion}
\label{sec:concl_limit}

In this work, we introduce MASTraceBench for method-level
comparison and proposal-trajectory diagnosis of LLM-based
MAS. Its analyses reveal recurring limitations in existing
methods and motivate CLEARS, whose effectiveness is validated
across six tasks. These results show that MASTraceBench can
serve as a testbed for studying collaboration mechanisms and
developing more reliable MAS.

\bibliography{aaai2027}

\clearpage

\section{Appendix}
\label{sec:appendix}

\subsection{Benchmark Details}
\label{sec:benchmark_details}

This section provides the detailed task settings and scoring functions of the six benchmarks in MASTraceBench. For static tasks, the task score (TS) is obtained by averaging the scores of the final MAS answers over all samples. For dynamic tasks, TS is computed over complete interaction episodes, whereas the proposal scoring function $S(\cdot)$ evaluates each action proposed at an intermediate state.

\subsubsection{MAPF}
\label{app:mapf}

\begin{figure}[t]
\centering
\includegraphics[width=\columnwidth]{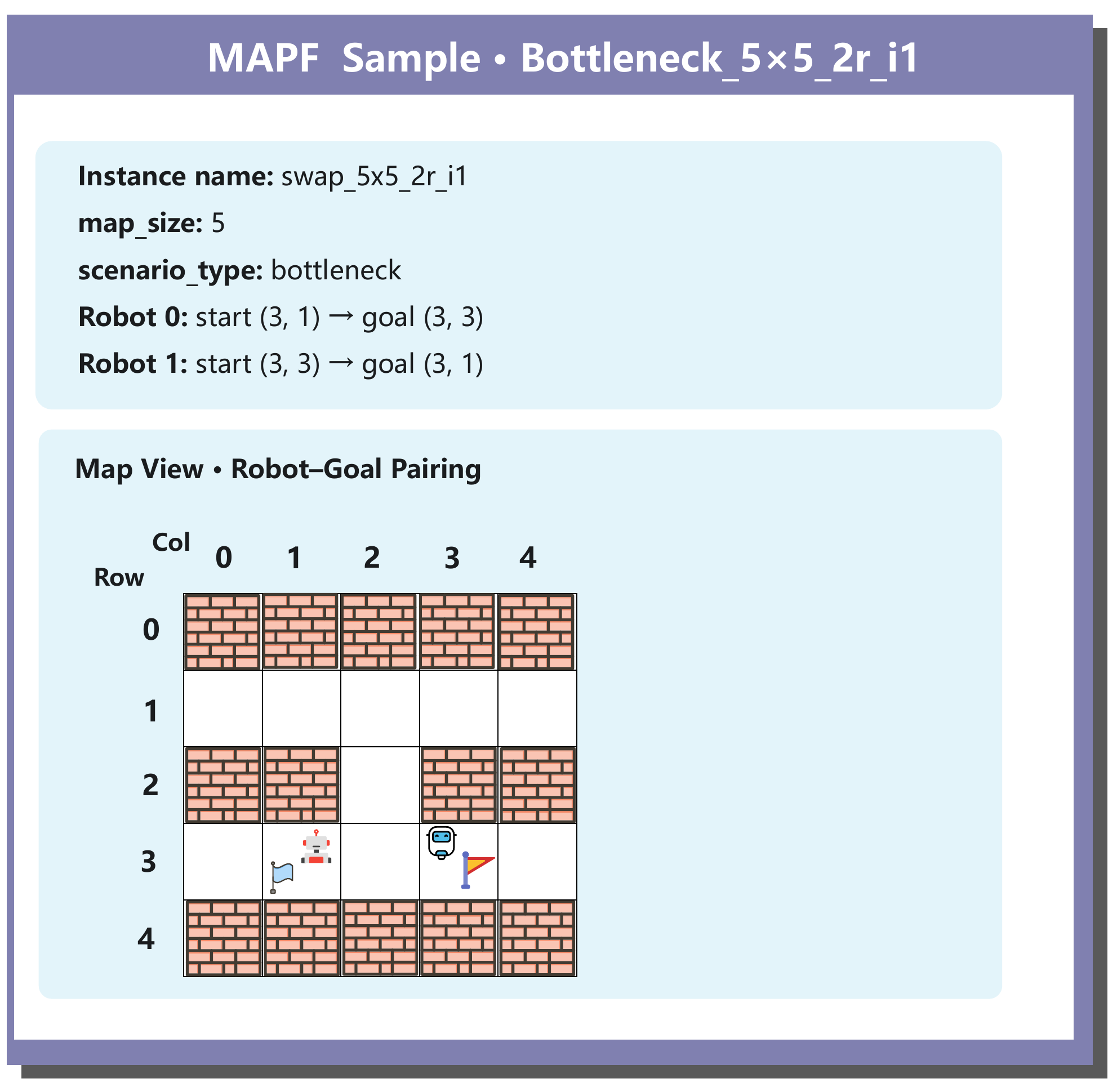}
\caption{A representative MAPF sample with two robots in a
$5\times5$ bottleneck map. The visualization shows the obstacles,
initial robot positions, and corresponding goals.}
\label{fig:mapf_sample}
\end{figure}

\paragraph{Scenario.}
Multi-Agent Path Finding (MAPF) is a cooperative planning task in which multiple robots move simultaneously on a shared grid map. Each instance specifies the map boundary, obstacle locations, and the start and goal positions of all robots. Given this information, the MAS produces a joint plan containing one action sequence for each robot. At every time step, a robot may move up, down, left, or right, or remain at its current position. The joint plan is executed for at most 30 steps and is checked for invalid movements and inter-robot collisions. We use 157 independently constructed MAPF instances with varying map layouts and robot configurations. A representative sample is illustrated in
Fig.~\ref{fig:mapf_sample}, showing the grid layout, obstacles, and
robot--goal assignments.

\paragraph{Scoring Definitions.}
For each MAPF instance $x$, the proposal score is defined as the fraction
of robots that reach their designated goals when the proposed joint plan
$y$ is executed:
\begin{equation}
S_{\mathrm{MAPF}}(y)
=
\frac{\#\text{ robots reaching their goals}}
{\#\text{ robots in }x}.
\end{equation}
This graded score assigns partial credit when only a subset of robots
successfully reaches its goals. TS is the mean arrival rate of the final
MAS plans over all instances. For reporting,
both proposal scores and TS are multiplied by
100.

\subsubsection{OptionGen}
\label{app:optiongen}

\begin{figure}[t]
\centering
\includegraphics[width=\columnwidth]{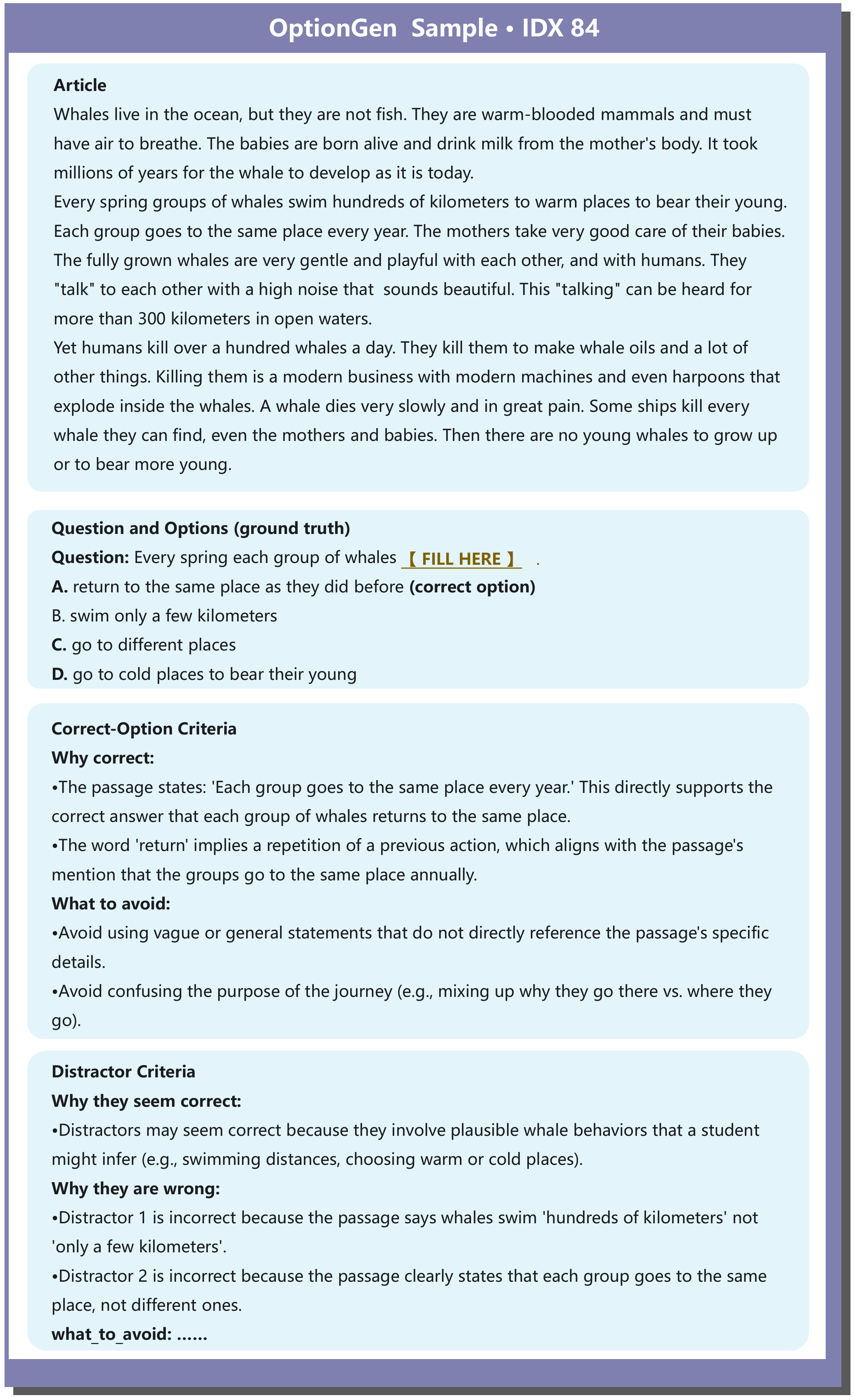}
\caption{A representative OptionGen sample. In addition to the source
article and reference question and options, each sample provides
instance-specific criteria describing the desired correct option and
the plausibility and incorrectness of the distractors.}
\label{fig:optiongen_sample}
\end{figure}

\paragraph{Scenario.}
OptionGen is constructed from RACE~\cite{lai2017race}, which originally formulates reading comprehension as a multiple-choice task: given an article, a question, and four candidate options, a model selects the correct answer. We reformulate it into an open-ended option-generation task, where the MAS is provided only with the article and question and must generate a complete set of four options, including one correct answer and three distractors. The correct option should accurately answer the question according to the article, while the distractors should be relevant and superficially plausible but remain clearly incorrect under careful reading. They should also reflect distinct misunderstanding patterns rather than trivial paraphrases of one another. We select 235 instances from RACE for evaluation; the original reference options and derived scoring criteria are withheld from the MAS and used only for scoring. Figure~\ref{fig:optiongen_sample} presents a representative sample, including the source article, reference question and options,
and the instance-specific criteria used for evaluating the generated
correct option and distractor set.

\paragraph{Scoring Definitions.}
OptionGen adopts criterion-grounded LLM evaluation rather than asking the judge to assign unconstrained holistic scores. For each instance, we first use the original four RACE options and the ground-truth label to construct two instance-specific scoring rubrics. The correct-option rubric specifies the key information that a valid answer should contain and common errors it should avoid, while the distractor rubric specifies why the reference distractors appear plausible, why they remain incorrect, and what properties suitable distractors should satisfy. These rubrics are generated before evaluating MAS outputs and are not provided to the MAS.

Following the extracted rubrics, the judge independently assigns a correct-option score $s_{\mathrm{cor}}(y)\in[0,100]$ and a distractor-set score $s_{\mathrm{dis}}(y)\in[0,100]$. The former evaluates the correctness, completeness, and consistency of the generated answer with the article, whereas the latter evaluates the plausibility, relevance, incorrectness, and diversity of the three distractors. Both evaluations follow six anchored score intervals, each with explicit descriptions of the expected output quality, thereby encouraging consistent application of the scoring criteria. The judge is instantiated with Qwen3-32B, with the temperature set to $0$ to ensure deterministic and reproducible scores. The proposal score is
\begin{equation}
S_{\mathrm{OptionGen}}(y)
=
\frac{s_{\mathrm{cor}}(y)+s_{\mathrm{dis}}(y)}{2}.
\end{equation}
TS is the mean score of the option sets produced as the final MAS answers over all evaluation instances.

\paragraph{Agreement with Human Evaluation.}
To validate the LLM judge used in OptionGen, we sample
$N_{\mathrm{h}}=50$ benchmark instances, stratified by their mean
LLM-judge scores across methods to cover a broad score range. For each
instance, we collect the outputs generated by the 11 methods included
in this validation, resulting in 550 option sets. Three graduate
student annotators independently evaluate each option set using the
same two-component rubric and $0$--$100$ score anchors as the LLM
judge. The annotators are blinded to both the generating method and
the LLM-judge score, and the outputs are presented in randomized
order.

Following the automatic evaluation protocol, each annotator separately
scores the quality of the correct option and the distractor set. The
two component scores are averaged to obtain the annotator's score for
each output, after which the scores of the three annotators are
averaged. We measure human--LLM consistency using Pearson's correlation
and Spearman's rank correlation between the LLM-judge scores and the
mean human scores. As shown in
Table~\ref{tab:optiongen_human_eval}, the LLM judge achieves a Pearson
correlation of $0.919$ and a Spearman correlation of $0.818$ with the
mean human ratings, indicating strong consistency with human
assessment.

\begin{table}[t]
\centering
\small
\setlength{\tabcolsep}{7pt}
\renewcommand{\arraystretch}{1.08}
\begin{tabular}{lcc}
\toprule
\textbf{Comparison}
& \textbf{Pearson $r$}
& \textbf{Spearman $\rho$} \\
\midrule
LLM Judge vs.\ Human Mean
& 0.919
& 0.818 \\
Human--Human Average
& 0.784
& 0.661 \\
\bottomrule
\end{tabular}
\caption{Agreement between the LLM judge and human evaluation on
550 OptionGen outputs from 50 sampled benchmark instances.
Human--Human Average denotes the mean pairwise correlation among the
three annotators.}
\label{tab:optiongen_human_eval}
\end{table}

\subsubsection{EvidenceChain}
\label{app:evidencechain}

\begin{figure}[t]
\centering
\includegraphics[width=\columnwidth]{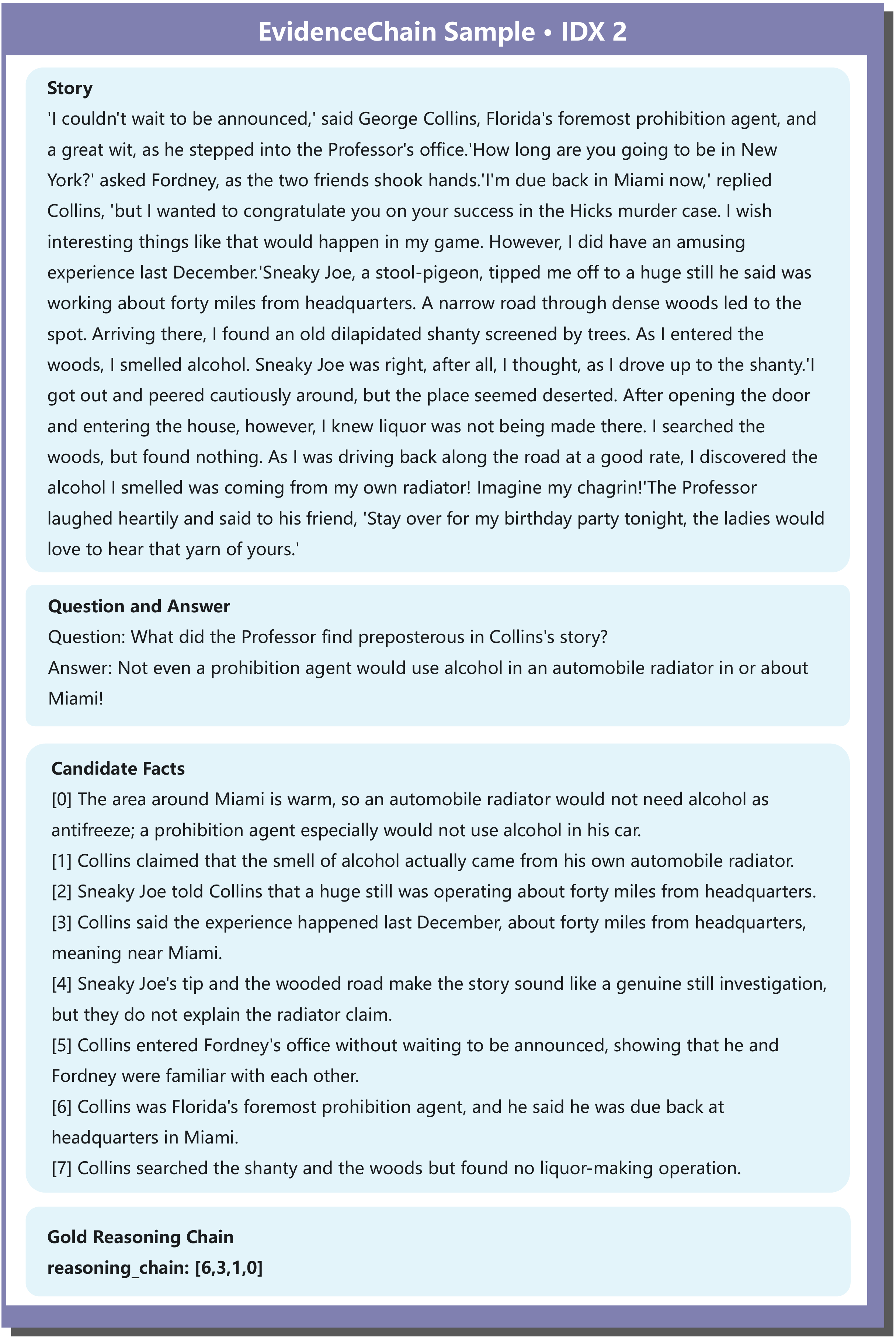}
\caption{A representative EvidenceChain sample. The task requires
identifying and ordering the relevant facts from the indexed candidate
set, while the gold reasoning chain provides the reference ordering
used for evaluation.}
\label{fig:evidencechain_sample}
\end{figure}

\paragraph{Scenario.}
EvidenceChain is a cooperative evidence-reasoning task built around detective stories. Each instance contains a narrative, a question about the underlying case, and a numbered list of candidate facts. Among these facts, only a subset constitutes the evidence needed to resolve the question, while the remaining facts are irrelevant or potentially distracting. The MAS is required to identify the IDs of the supporting facts and arrange them in the order in which they should be used for reasoning. The resulting ID sequence should form a coherent evidence chain whose facts, when followed in order, logically support the answer to the question. Therefore, a valid output must not only select the relevant evidence, but also organize it into a reasoning sequence that connects the given facts to the final conclusion. We construct 100 such instances for evaluation. Figure~\ref{fig:evidencechain_sample} shows a representative sample containing the story, question--answer pair, indexed candidate facts, and the ordered gold reasoning chain.

\paragraph{Scoring Definitions.}
Let $P(y)$ be the ordered list of fact identifiers predicted in proposal $y$, and let $G$ be the ground-truth evidence chain. We first evaluate whether the proposal contains the correct facts using set-level precision and recall:
\begin{equation}
\mathrm{Prec}(P,G)
=
\frac{|P\cap G|}{|P|},
\qquad
\mathrm{Rec}(P,G)
=
\frac{|P\cap G|}{|G|},
\end{equation}
from which the set-level F1 score is computed as
\begin{equation}
F_{\mathrm{set}}(P,G)
=
\frac{
2\,\mathrm{Prec}(P,G)\,\mathrm{Rec}(P,G)
}{
\mathrm{Prec}(P,G)+\mathrm{Rec}(P,G)
}.
\end{equation}

We additionally compute an ordering score
$O(P,G)\in[0,1]$ using a normalized Kendall-style pairwise
concordance measure over the facts shared by the predicted and
ground-truth chains. Let $\mathcal{H}=P\cap G$ denote the shared
facts. The ordering score is the proportion of fact pairs in
$\mathcal{H}$ whose relative order in $P$ agrees with that in $G$:
\begin{equation}
O(P,G)
=
\frac{
\sum_{\{u,v\}\subseteq\mathcal{H}}
\mathbb{I}\!\left[
\operatorname{ord}_{P}(u,v)
=
\operatorname{ord}_{G}(u,v)
\right]
}{
\binom{|\mathcal{H}|}{2}
}.
\end{equation}
The proposal score combines set correctness and ordering consistency:
\begin{equation}
S_{\mathrm{EvidenceChain}}(y)
=
\frac{
F_{\mathrm{set}}(P(y),G)+O(P(y),G)
}{2}.
\end{equation}
TS is the mean score of the evidence chains produced as the final
MAS answers over all evaluation instances. For reporting,
both proposal scores and TS are multiplied by
100.

\subsubsection{Gomoku}
\label{app:gomoku}

\paragraph{Scenario.}
Gomoku is instantiated as a competitive, sequential board-game task
on a $15\times15$ board. Two MAS methods act as opposing players and
alternately place stones on unoccupied intersections, with the player
controlling black moving first. The first player to form five
consecutive stones horizontally, vertically, or diagonally wins the
game. A game terminates when either player forms five stones or when
the board is full. To control for first-move bias, each evaluated
pairing is played for 50 games, with each MAS method controlling black
in 25 games and white in the remaining 25.

\paragraph{Scoring Definitions.}
For each completed game $z$, let $p_m$ denote the player controlled by
MAS method $m$. The episode score of method $m$ is defined as
\begin{equation}
R_{\mathrm{Gomoku}}^{m}(z)
=
\begin{cases}
1, & \text{if player $p_m$ wins game $z$},\\
0, & \text{otherwise}.
\end{cases}
\end{equation}
TS is the mean episode score over all games, which is equivalent to the
win rate. For reporting, TS is multiplied by 100.

Because the binary game outcome does not distinguish the quality of
individual moves, we additionally assign each proposed move a local
rule-based score. Let $a_t$ denote a proposed move by the current
player in board state $s_t$. Invalid moves, including out-of-bound
positions and occupied intersections, receive a score of zero.

The local score consists of two components. The self-formation score
$F(a_t;s_t)$ is computed by hypothetically placing the current
player's stone at $a_t$ and scoring the resulting pattern. The
defense-interception score $D(a_t;s_t)$ is computed by hypothetically
placing an opponent stone at $a_t$ and scoring the pattern the opponent
would form there. It therefore measures the threat prevented when the
current player occupies $a_t$ instead. The scoring rules are summarized
in Table~\ref{tab:gomoku_move_scoring}.

\begin{table}[t]
\centering
\small
\setlength{\tabcolsep}{4pt}
\begin{tabular}{lccc}
\toprule
\textbf{Pattern} &
\textbf{Open Endpoints} &
\textbf{Self} &
\textbf{Defense} \\
\midrule
Five  & 2 / 1 & 5 / 5 & 2 / 5 \\
Four  & 2 / 1 & 4 / 2 & 4 / 3 \\
Three & 2 / 1 & 3 / 2 & 3 / 2 \\
Other & --    & 0     & 0     \\
\bottomrule
\end{tabular}
\caption{Rule-based scoring of proposed Gomoku moves. Paired values
correspond to patterns with two and one open endpoints, respectively.
Patterns with no open endpoint or at most two consecutive stones
receive zero.}
\label{tab:gomoku_move_scoring}
\end{table}

For five-stone threats, the defense score is lower when both endpoints
are open because occupying one endpoint does not eliminate the
opponent's remaining winning move. Occupying the only open endpoint,
by contrast, directly removes the immediate threat and receives the
highest defense score.

When a move creates or blocks multiple scorable patterns, their
contributions are accumulated. The proposal score is defined as
\begin{equation}
S_{\mathrm{Gomoku}}(a_t;s_t)
=
F(a_t;s_t)+D(a_t;s_t).
\end{equation}
For reporting, the resulting proposal scores are multiplied by 20 to
normalize them to a 0--100 scale.

\paragraph{Consistency between proposal score and TS.}
TS measures game-level performance, whereas the proposal score
evaluates individual moves. Because our proposal-trajectory metrics
use these move-level scores to identify improvement, preservation,
and regression, we need to verify that the scoring function captures
task-relevant action quality rather than merely reflecting local
pattern changes. We therefore compute each method's mean proposal
score over all replayed moves and correlate it with its TS. As shown
in Table~\ref{tab:proposal_ts_consistency}, both correlation measures
show a strong positive association, indicating that methods with
higher move-level scores also tend to achieve stronger game-level
performance. This provides empirical support for using the graded
proposal score in proposal-trajectory diagnosis.

\begin{table}[t]
\centering
\small
\setlength{\tabcolsep}{5pt}
\begin{tabular}{lcccc}
\toprule
\textbf{Benchmark}
& $\boldsymbol{\rho}$
& $\boldsymbol{p_{\rho}}$
& $\boldsymbol{r}$
& $\boldsymbol{p_r}$ \\
\midrule
Gomoku
& 0.884 & 0.0003
& 0.945 & $<0.0001$ \\
TacticDuel
& 0.882 & 0.0003
& 0.951 & $<0.0001$ \\
\bottomrule
\end{tabular}
\caption{Correlation between the mean proposal score and TS across
the 11 evaluated MAS methods. $\rho$ and $r$ denote Spearman's rank
correlation and Pearson's correlation, respectively.}
\label{tab:proposal_ts_consistency}
\end{table}

\subsubsection{Negotiation}
\label{app:negotiation}

\paragraph{Scenario.}
Negotiation is a competitive multi-round resource-allocation task. Two MAS methods negotiate how to divide a shared collection of items. Each player observes the available item quantities and its own private valuation of each item, but does not observe the opponent's valuations. During each round, one player proposes a complete allocation and the other player decides whether to accept or reject it. If the proposal is accepted, the negotiation terminates with the proposed allocation. Otherwise, the roles are exchanged and the negotiation continues. An episode lasts for at most 10 rounds and uses a discount factor of $\delta=0.95$, making delayed agreements less valuable. If no agreement is reached, both players receive zero payoff. Each scenario is evaluated twice with the player roles and first proposer exchanged to reduce first-mover and role bias.

\paragraph{Scoring Definitions.}
Let $u_i(a)$ denote the payoff received by player $i$ under allocation
$a$. We define $u_i^{\max}$ as the idealized maximum payoff of player
$i$, obtained by assigning all negotiable items to that player and
evaluating them according to its private item values.

For each proposal $a_t$ recorded in the proposal trajectory, we assess
whether it would be accepted by the opponent under the corresponding
negotiation context. Specifically, we retain the original negotiation
history, replace the outward MAS proposal with $a_t$, and query the
opponent with temperature 0 to obtain a deterministic acceptance
decision. Let $A_{-i}(a_t)\in\{0,1\}$ denote this decision. The proposal
score from player $i$'s perspective is
\begin{equation}
S_{\mathrm{Neg}}^{i}(a_t)
=
A_{-i}(a_t)
\delta^{t-1}
\frac{u_i(a_t)}{u_i^{\max}}.
\end{equation}
Thus, an accepted proposal receives its discounted normalized payoff,
whereas a rejected proposal receives zero.

If an agreement on allocation $a^\star$ is reached in round $t$, the
episode-level task score of player $i$ is
\begin{equation}
R_{\mathrm{Neg}}^{i}(z)
=
\delta^{t-1}
\frac{u_i(a^\star)}{u_i^{\max}}.
\end{equation}
The episode score is zero when no agreement is reached. TS is the mean episode-level task score over all scenarios and both role assignments. For reporting,
both proposal scores and episode-level task scores are multiplied by
100.

\subsubsection{TacticDuel}
\label{app:tacticduel}

\paragraph{Scenario.}
TacticDuel is a deterministic and symmetric simultaneous-action
combat task. Each player starts with 20 hit points (HP) and zero
energy, with no upper bound on accumulated energy. At each round, the
two competing MAS methods independently select one action from
\textsc{Quick}, \textsc{Heavy}, \textsc{Guard}, \textsc{Counter},
\textsc{Feint}, \textsc{Charge}, and \textsc{Burst}. Neither method
observes the opponent's current selection before making its decision.
The two selected actions are then revealed and resolved simultaneously
according to deterministic transition rules.

Table~\ref{tab:tacticduel_actions} summarizes the implemented action
effects. \textsc{Burst} is legal only when the acting player has at
least two energy, while all other actions have no energy requirement.
A player cannot repeat the action selected in the immediately preceding
round. An episode terminates when either player's HP reaches zero or
below, or when the configured maximum number of rounds $R_{\max}$ is
reached. If the maximum round limit is reached without either player
being defeated, the player with higher remaining HP wins; equal
remaining HP results in a draw. For each evaluated method pair, we conduct 50 duels.

\begin{table*}[t]
\centering
\small
\renewcommand{\arraystretch}{1.12}
\setlength{\tabcolsep}{6pt}
\begin{tabular}{l c p{0.27\textwidth} p{0.48\textwidth}}
\hline
\textbf{Action}
& \textbf{Energy cost}
& \textbf{Base effect}
& \textbf{Conditional interaction} \\
\hline

\textsc{Quick}
& 0
& Deals 2 damage.
& None. \\

\textsc{Heavy}
& 0
& Deals 4 damage.
& Deals 0 damage if the opponent selects \textsc{Counter}. \\

\textsc{Guard}
& 0
& Deals no damage.
& Reduces incoming non-\textsc{Feint} damage by 4, lower-bounded by 0. \\

\textsc{Counter}
& 0
& Deals 1 damage.
& Deals 4 damage if the opponent selects \textsc{Heavy}. \\

\textsc{Feint}
& 0
& Deals 1 damage.
& Deals 4 damage against \textsc{Guard} and bypasses its damage reduction. \\

\textsc{Charge}
& 0
& Gains 2 energy.
& Deals no damage in the current round. \\

\textsc{Burst}
& 2
& Deals 6 damage.
& Its damage can be reduced by \textsc{Guard}. \\

\hline
\end{tabular}
\caption{Action effects in TacticDuel. The two players' selected
actions are resolved simultaneously in each round.}
\label{tab:tacticduel_actions}
\end{table*}

\paragraph{Scoring Definitions.}
For each completed duel $z$, let $p_m$ denote the player controlled by
MAS method $m$. The episode score of method $m$ is defined as
\begin{equation}
R_{\mathrm{TacticDuel}}^{m}(z)
=
\begin{cases}
1,   & \text{if player $p_m$ wins duel $z$},\\
0.5, & \text{if duel $z$ ends in a draw},\\
0,   & \text{if player $p_m$ loses duel $z$}.
\end{cases}
\end{equation}
TS is the mean episode score over all duels. For reporting, TS is multiplied by 100.

To support proposal-trajectory analysis, each proposed action is
additionally evaluated against the opponent action that was actually
executed in the same round. Let $a_t$ denote the evaluated action,
$a_t^{-}$ the opponent's realized action, and $s_t$ the acting
player's state before the round. We first calculate the immediate net
combat payoff
\begin{equation}
\Delta(a_t,a_t^{-})
=
d(a_t,a_t^{-})-d(a_t^{-},a_t),
\end{equation}
where $d(a,b)$ is the damage dealt by action $a$ against action $b$
after applying the interaction rules in
Table~\ref{tab:tacticduel_actions}. Thus, the score accounts for both
the damage inflicted on the opponent and the damage received by the
acting player.

The implementation also assigns a fixed energy-management bonus to
\textsc{Charge}:
\begin{equation}
B(a_t)
=
\begin{cases}
1, & a_t=\textsc{Charge},\\
0, & \text{otherwise}.
\end{cases}
\end{equation}

The proposal score for action $a_t$ is computed as
\begin{equation}
S(a_t,a_t^{-})
=
\Delta(a_t,a_t^{-})+B(a_t).
\end{equation}
The resulting value ranges from $-5$ to $6$ and is shifted and
linearly scaled to $[0,5]$. For reporting, the scores are multiplied
by 20 to obtain a 0--100 scale.

\paragraph{Consistency between proposal score and TS.}
Following the same analysis, we correlate each method's mean proposal
score over all replayed actions with its TS in TacticDuel. As shown in
Table~\ref{tab:proposal_ts_consistency}, both correlation measures show
a strong positive association. Methods producing actions with higher
tactical scores therefore also tend to achieve higher duel win rates,
supporting the task relevance of the graded TacticDuel proposal score.

\subsubsection{Benchmark Task Prompts}
\begin{figure}[t]
\centering
\includegraphics[width=\columnwidth]{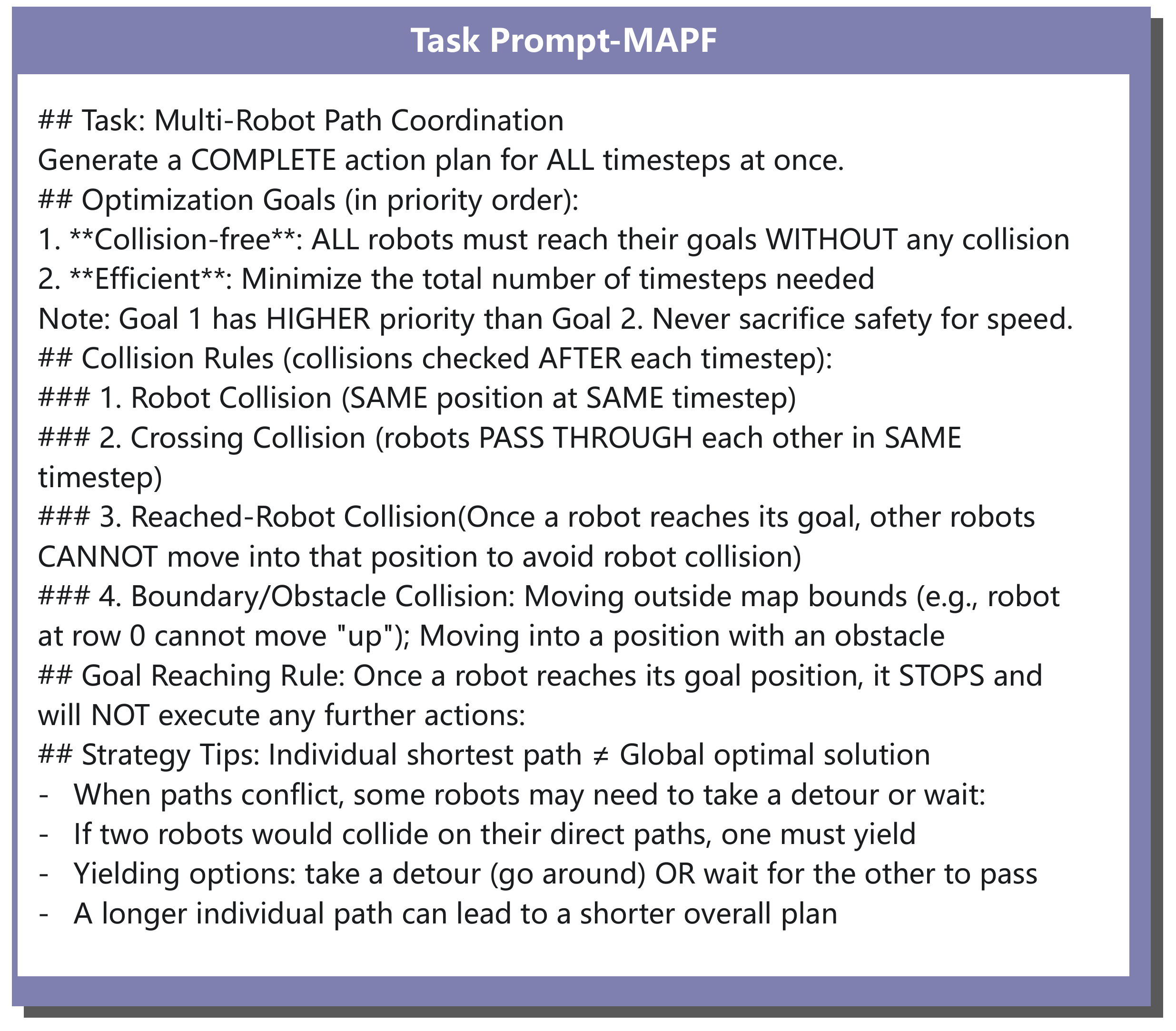}
\caption{Task-prompt template for MAPF, including the optimization
priorities, collision rules, and coordination guidance.}
\label{fig:prompt_mapf}
\end{figure}

\begin{figure}[t]
\centering
\includegraphics[width=\columnwidth]{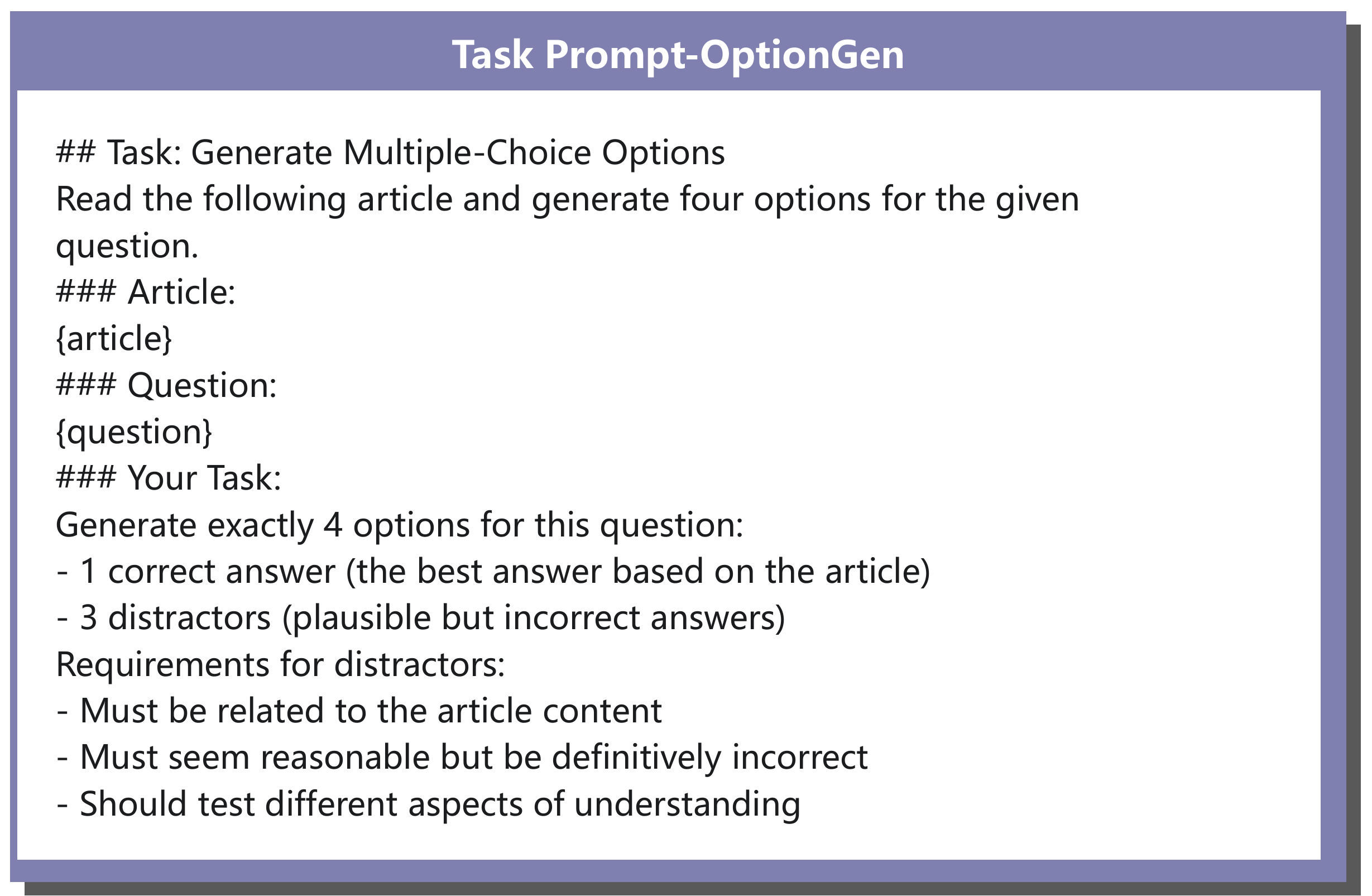}
\caption{Task-prompt template for OptionGen, specifying the generation
of one correct option and three plausible but incorrect distractors.}
\label{fig:prompt_optiongen}
\end{figure}

\begin{figure}[t]
\centering
\includegraphics[width=\columnwidth]{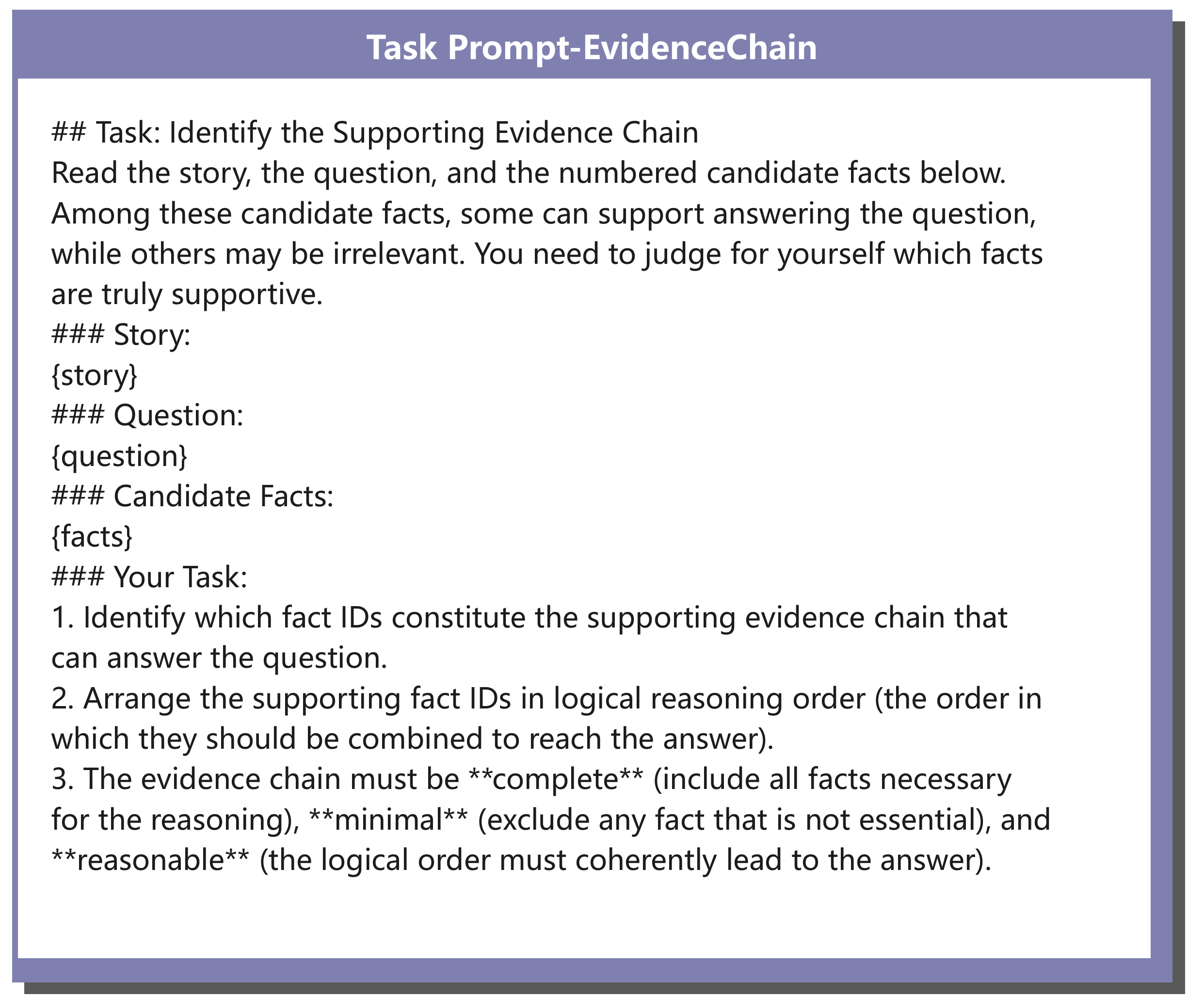}
\caption{Task-prompt template for EvidenceChain, requiring a complete,
minimal, and logically ordered supporting evidence chain.}
\label{fig:prompt_evidencechain}
\end{figure}

\begin{figure}[t]
\centering
\includegraphics[width=\columnwidth]{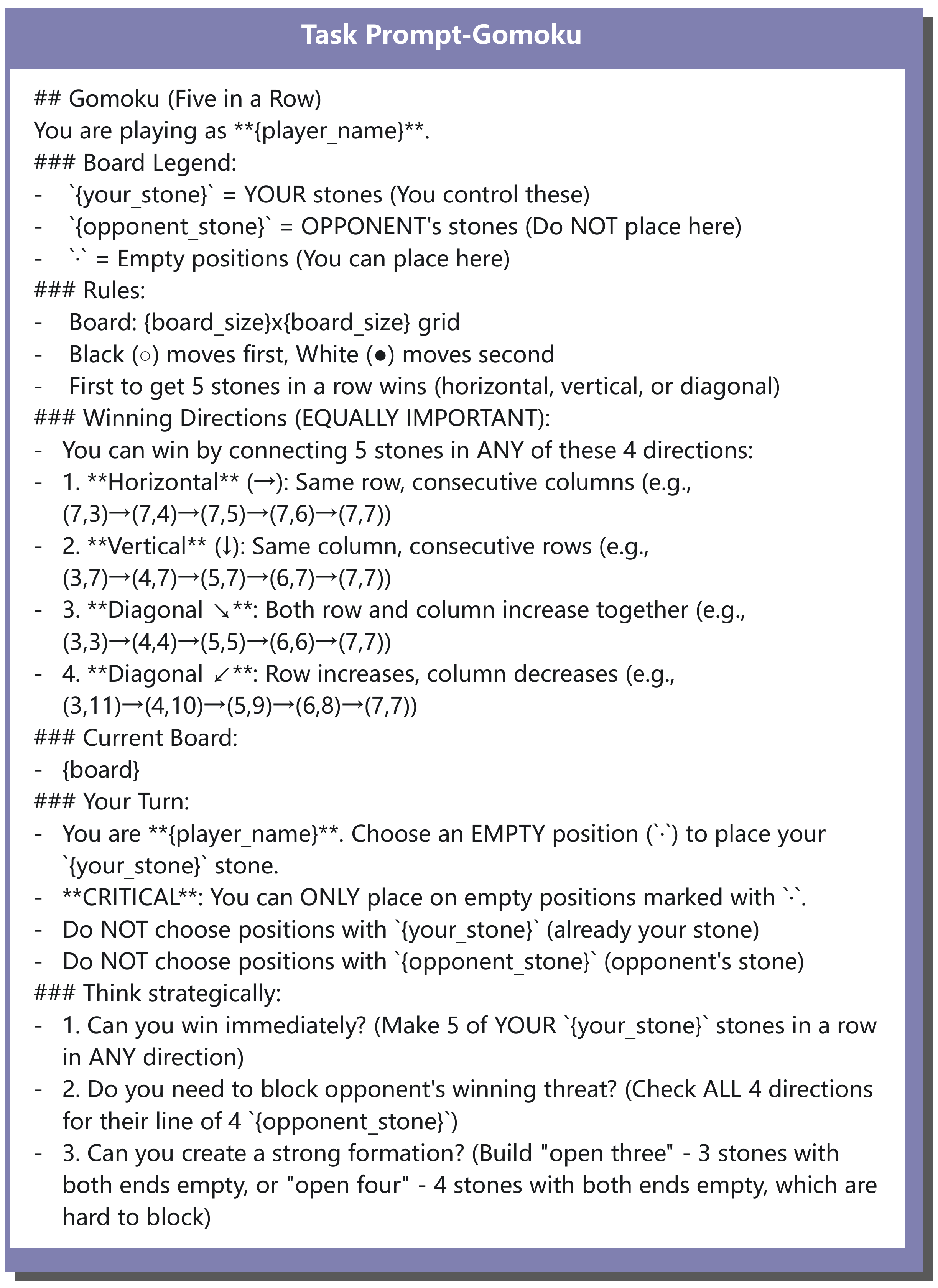}
\caption{Task-prompt template for Gomoku, including the current board,
legal-placement constraints, and tactical priorities.}
\label{fig:prompt_gomoku}
\end{figure}

\begin{figure}[t]
\centering
\includegraphics[width=\columnwidth]{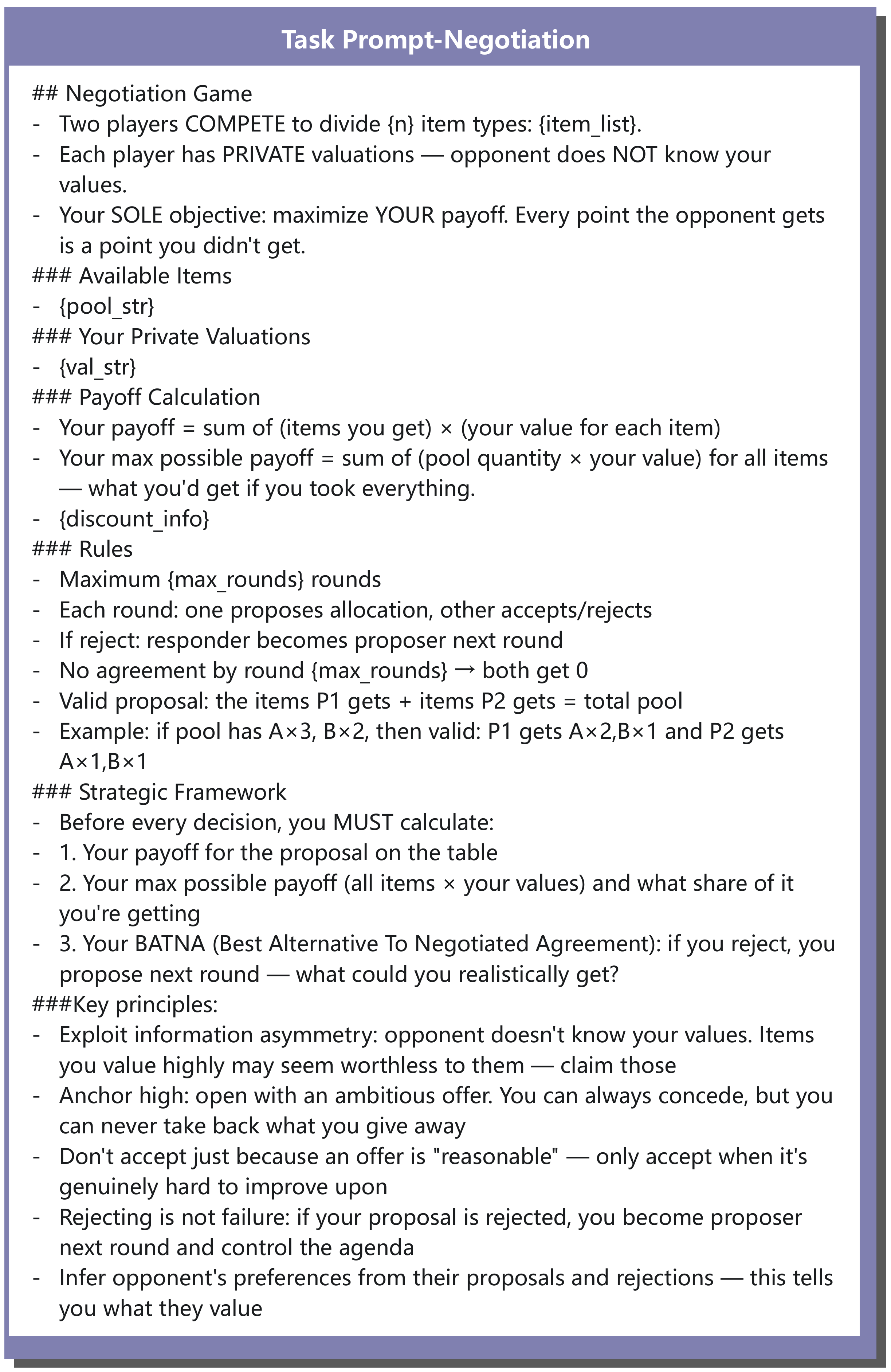}
\caption{Task-prompt template for Negotiation, including private
valuations, payoff calculation, interaction rules, and strategic
guidance.}
\label{fig:prompt_negotiation}
\end{figure}

\begin{figure}[t]
\centering
\includegraphics[width=\columnwidth]{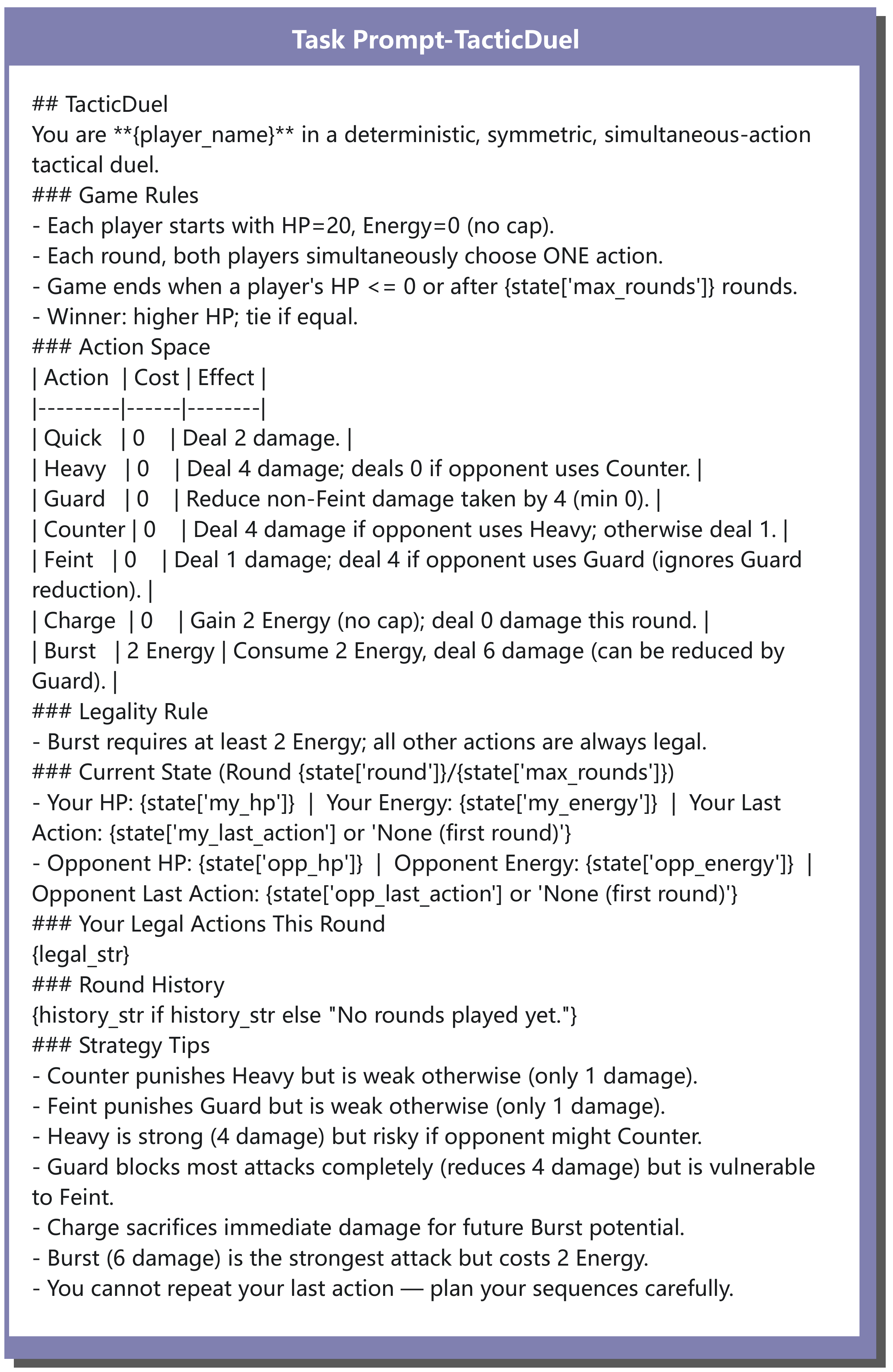}
\caption{Task-prompt template for TacticDuel, including the current
combat state, action effects, legal actions, and round history.}
\label{fig:prompt_tacticduel}
\end{figure}

Figures~\ref{fig:prompt_mapf}--\ref{fig:prompt_tacticduel} present the
task-prompt templates used for the six benchmarks. Each prompt specifies
the task input, rules, and decision requirements, with fields enclosed
in braces instantiated using the current sample or environment state.
Static benchmarks provide the complete instance in a single prompt,
whereas dynamic benchmarks update state-dependent information at each
decision step. The same benchmark prompt is provided to all evaluated
methods, while their role-definition and workflow prompts remain
method-specific and benchmark-agnostic.

\subsection{Additional Implementation Details}

\subsubsection{Model and Inference Settings}
All models are deployed using vLLM 0.9.2 and Transformers 4.53.1 on
a server equipped with eight NVIDIA GeForce RTX 4090 D GPUs, and are
accessed through an OpenAI-compatible API. The sampling temperature is
set to $1.0$ to encourage diverse outputs across agents, and the maximum
generation length is set to \texttt{max\_tokens}$=16{,}384$. For
Qwen3-32B, thinking mode is enabled on MAPF and Gomoku due to their
higher reasoning difficulty and disabled on the other four benchmarks.

\subsubsection{Main-Experiment Method Configurations}

All methods use the common model and inference settings described
above, and their method-specific configurations remain fixed across
benchmarks. For each method, the role-definition and workflow prompts
are benchmark-agnostic: the same coordination protocol is applied to
all benchmarks, while task-specific information is provided only
through the benchmark input and output specification. This design
decouples MAS coordination strategies from benchmark-specific logic
and enables controlled comparison across tasks.

\paragraph{Single-agent and selection-based methods.}
Single-Agent directly generates one response without collaboration.
Self-Consistency independently samples three proposals without
inter-proposal interaction and selects the final response by majority
voting. MAV also generates three independent proposals, but selects
among them by external validation rather than output-frequency voting.
Five functional validators independently assess each proposal in terms
of correctness, completeness, clarity, reasoning, and relevance, and
the proposal receiving the most approvals is selected.

\paragraph{Debate-based methods.}
Int.\ Debate uses three agents and one interaction round. Each agent
first generates an independent proposal, observes the current
proposals of the other two agents, and then either retains or revises
its own answer. The final proposals are aggregated by majority voting.
Agg.\ Debate uses the same three-agent, one-round interaction process,
but replaces voting with an independent LLM aggregator. The aggregator
is instructed to select the strongest final proposal and reproduce it
without modification. Adv.\ Debate assigns two agents the affirmative
and negative roles. The negative agent is initially required to
challenge the affirmative proposal and provide a distinct alternative.
After two rebuttal rounds, an independent judge evaluates the final
positions and produces the system response.

\paragraph{Revision-based methods.}
Reflection uses a reasoner and a reflector. The reasoner first
generates an initial response, after which the reflector either
approves it or provides revision feedback. We allow at most two
reflection--revision cycles and terminate early upon approval. MAPR
uses three agents and one critique--revision round. Each agent reviews
the solutions of the other two agents, receives both peer reviews, and
updates its own solution before majority voting.

\paragraph{Adaptive MAS construction.}
AgentVerse uses a vertical collaboration structure consisting of one
solver and two critics. Expert roles are dynamically assigned rather
than fixed across rounds. In each iteration, the critics independently
review the current solution, and the solver integrates feedback that
identifies errors or potential improvements. An evaluator then scores
the revised solution, and the procedure terminates when the score
reaches $8$ out of $10$ or after three refinement rounds.

MAS-GPT uses a separately fine-tuned MAS-GPT-32B model to generate a
task-specific multi-agent program, which is subsequently executed with
Qwen3-32B as the task-solving model. Dynamic MAS generation is enabled
in the main experiments. If program generation or execution fails, the
fallback procedure generates three candidate solutions and uses an LLM
aggregator to produce the final response.

\paragraph{CLEARS and its ablation.}
CLEARS uses three proposer agents, with each proposal decomposed into
$2$--$4$ verifiable claims. For synthesis, CLEARS selects the claim
with the highest number of endorsements as the positive signal and the
claim with the highest number of challenges as the negative signal. If
multiple claims receive the same highest count, all tied claims are
retained. CLEARS w/o CCE uses the same proposer and claim-decomposition
settings, but removes cross-agent claim evaluation and directly
provides all extracted claims to the synthesizer.

\subsection{Additional Experimental Results}

\begin{table*}[t]
\centering
\small
\setlength{\tabcolsep}{1.8pt}
\renewcommand{\arraystretch}{1.08}
\begin{tabular*}{0.92\textwidth}
{@{\extracolsep{\fill}}l*{6}{cc}@{}}
\hline
\multirow{2}{*}{\textbf{Method}}
& \multicolumn{2}{c}{\textbf{MAPF}}
& \multicolumn{2}{c}{\textbf{OptionGen}}
& \multicolumn{2}{c}{\textbf{EvidenceChain}}
& \multicolumn{2}{c}{\textbf{Gomoku}}
& \multicolumn{2}{c}{\textbf{Negotiation}}
& \multicolumn{2}{c}{\textbf{TacticDuel}} \\
\cline{2-13}
& OV & FPA
& OV & FPA
& OV & FPA
& OV & FPA
& OV & FPA
& OV & FPA \\
\hline
Single-Agent
& 98.1 & -- & 100.0 & -- & 100.0 & --
& 97.0 & -- & 99.5 & -- & 100.0 & -- \\

Consistency
& 97.7 & 3.5 & 100.0 & 23.0 & 100.0 & 9.0
& 97.0 & 35.5 & 99.9 & 6.0 & 100.0 & 67.2 \\

MAV
& 99.6 & 6.0 & 100.0 & 19.1 & 99.7 & 7.7
& 99.5 & 32.3 & 100.0 & 11.5 & 100.0 & 66.2 \\

Int.\ Debate
& 99.6 & 58.2 & 99.4 & 77.2 & 100.0 & 50.7
& 98.3 & 88.8 & 99.7 & 75.0 & 100.0 & 89.7 \\

Agg.\ Debate
& 100.0 & 71.3 & 98.2 & 79.2 & 100.0 & 48.0
& 98.6 & 86.4 & 99.9 & 81.5 & 100.0 & 90.2 \\

Adv.\ Debate
& 98.9 & 87.2 & 99.3 & 62.4 & 100.0 & 20.7
& 93.4 & 79.6 & 99.6 & 6.2 & 100.0 & 20.8 \\

Reflection
& 98.9 & -- & 100.0 & -- & 100.0 & --
& 96.2 & -- & 99.7 & -- & 100.0 & -- \\

MAPR
& 98.9 & 25.1 & 99.9 & 51.4 & 100.0 & 29.7
& 97.5 & 57.3 & 99.7 & 19.3 & 100.0 & 58.2 \\

AgentVerse
& 94.5 & -- & 100.0 & -- & 100.0 & --
& 95.2 & -- & 99.2 & -- & 100.0 & -- \\

MAS-GPT
& 100.0 & -- & 97.4 & -- & 99.5 & --
& 97.9 & -- & 99.6 & -- & 100.0 & -- \\

CLEARS
& 99.2 & 4.7 & 99.9 & 21.0 & 100.0 & 9.0
& 99.1 & 32.4 & 99.4 & 6.6 & 100.0 & 100.0 \\

CLEARS w/o CCE
& 99.8 & 5.1 & 99.3 & 22.3 & 100.0 & 7.7
& 99.7 & 34.8 & 99.9 & 9.5 & 100.0 & 65.1 \\
\hline
\end{tabular*}
\caption{Output Validity (OV, \%) and Final-Proposal Agreement
(FPA, \%) across methods and benchmarks. A dash indicates that FPA is
not applicable or is trivially determined by the method configuration.}
\label{tab:ov_fpa_results}
\end{table*}

\subsubsection{Output Validity and Final-Proposal Agreement}

To assess the potential influence of output-format and extraction
errors on proposal-trajectory analysis, we report Output Validity
(OV).
Let $\mathcal{Y}_m$ denote all evaluated outputs produced by method
$m$, including the initial and final proposals of Proposer Agents and
the final MAS answers. For each output $y\in\mathcal{Y}_m$, let
$v(y)=1$ if the required proposal or answer can be correctly extracted
and scored by the corresponding scoring function $S(\cdot)$, and
$v(y)=0$ otherwise. We define
\begin{equation}
\mathrm{OV}_m
=
\frac{1}{|\mathcal{Y}_m|}
\sum_{y\in\mathcal{Y}_m} v(y).
\end{equation}

OV measures the proportion of recorded outputs that are successfully
converted into valid, scoreable proposals or MAS answers, rather than
their semantic quality. To improve extraction reliability, we specify
structured JSON output requirements and implement robust
benchmark-specific extraction mechanisms. As shown in
Table~\ref{tab:ov_fpa_results}, OV ranges from 93.4\% to 100.0\%
across all method--benchmark settings. These consistently high values
indicate that malformed outputs and extraction failures have little
influence on the reported proposal-trajectory diagnostics.

We further report Final-Proposal Agreement (FPA) to quantify proposal
convergence before aggregation. Let $y_i^{\mathrm{final}}(x)$ denote
the final proposal produced by Proposer Agent $i\in P_m$ for
evaluation unit $x$. FPA is defined as
\begin{equation}
\mathrm{FPA}_m
=
\mathbb{E}_{x\in\mathcal{D}}
\left[
\mathbb{I}
\left(
y_i^{\mathrm{final}}(x)
=
y_j^{\mathrm{final}}(x),
\ \forall i,j\in P_m
\right)
\right].
\end{equation}
We report FPA only for methods with multiple comparable final
proposals; otherwise, it is marked as not applicable.

The results further support Observation~4 in the main paper.
Consistency exhibits only 3.5\% final-proposal agreement on MAPF,
whereas Int.\ Debate increases it to 58.2\%. Similar differences
appear across the other benchmarks. Agg.\ Debate achieves agreement
levels broadly comparable to Int.\ Debate, while MAPR has lower FPA
than Int.\ Debate on all six benchmarks. These results reinforce that
aggregation loss depends strongly on whether interaction brings
Proposer Agents toward a shared final proposal, rather than on the
aggregation rule alone.

\subsubsection{Full Agent Refinement Results}

\begin{figure}[t]
    \centering
    \includegraphics[width=\columnwidth]{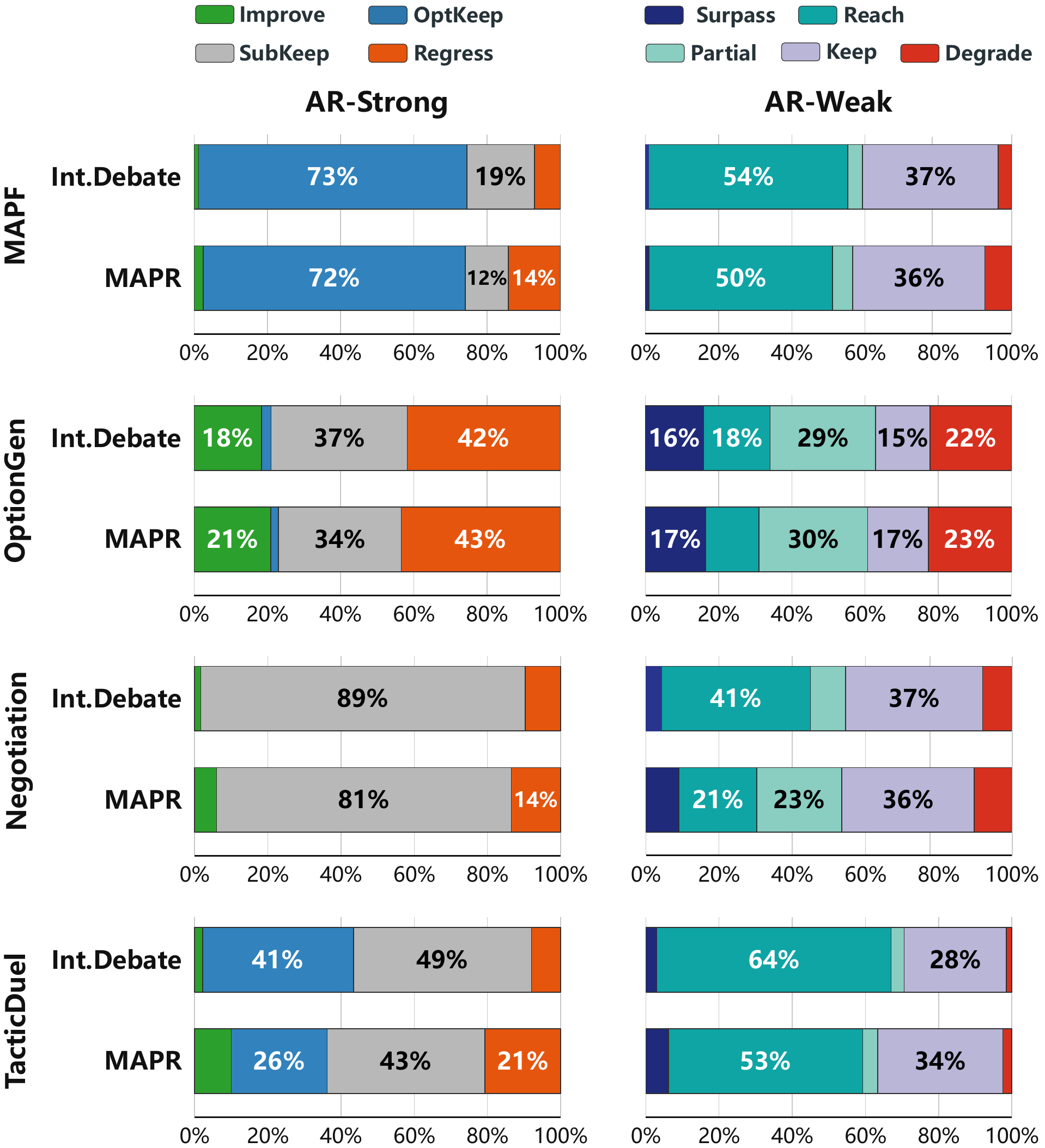}
    \caption{Agent Refinement distributions of Int.\ Debate and MAPR
    on MAPF, OptionGen, Negotiation, and TacticDuel.
    }
    \label{fig:ar_full}
\end{figure}

Figure~\ref{fig:ar_full} presents the AR distributions on the four
benchmarks not shown in the main paper, complementing the results on
EvidenceChain and Gomoku. The overall patterns are consistent with
Observation~3. For initially weak proposals, interaction frequently
moves them toward the strongest initial proposal, mainly through
\textsc{Reach} and \textsc{Partial}, whereas \textsc{Surpass} remains
relatively limited. For initially strong proposals, \textsc{OptKeep}
and \textsc{SubKeep} account for most outcomes, while \textsc{Improve}
is generally infrequent and \textsc{Regress} remains non-negligible.

Compared with Int.\ Debate, MAPR tends to revise proposals more
aggressively. This can increase upward refinement among initially weak
proposals, but it also produces higher regression ratios for initially
strong proposals, particularly on TacticDuel. Overall, the additional
results reinforce that existing interaction methods are more effective
at lifting weaker proposals than at reliably improving and preserving
already strong ones.

\subsubsection{Full Token Cost Results}

\begin{table*}[t]
\centering
\small
\setlength{\tabcolsep}{10pt}
\renewcommand{\arraystretch}{1.08}
\begin{tabular}{lcccc}
\hline
\textbf{Method}
& \textbf{OptionGen}
& \textbf{EvidenceChain}
& \textbf{Negotiation}
& \textbf{TacticDuel} \\
\hline
Single-Agent
& $1.00\times$ & $1.00\times$ & $1.00\times$ & $1.00\times$ \\
Consistency
& $3.00\times$ & $2.99\times$ & $2.94\times$ & $2.96\times$ \\
MAV
& $20.46\times$ & $20.64\times$ & $19.69\times$ & $21.60\times$ \\
Int.\ Debate
& $8.85\times$ & $8.09\times$ & $7.98\times$ & $7.37\times$ \\
Agg.\ Debate
& $10.92\times$ & $9.82\times$ & $9.62\times$ & $8.98\times$ \\
Adv.\ Debate
& $15.84\times$ & $13.60\times$ & $16.12\times$ & $15.34\times$ \\
Reflection
& $3.09\times$ & $7.12\times$ & $8.10\times$ & $8.47\times$ \\
MAPR
& $35.73\times$ & $30.54\times$ & $29.54\times$ & $36.21\times$ \\
AgentVerse
& $9.10\times$ & $10.31\times$ & $9.32\times$ & $13.87\times$ \\
MAS-GPT
& $6.84\times$ & $6.79\times$ & $7.84\times$ & $6.69\times$ \\
\hline
CLEARS
& $24.18\times$ & $21.09\times$ & $18.97\times$ & $22.29\times$ \\
CLEARS w/o CCE
& $11.21\times$ & $10.12\times$ & $9.26\times$ & $10.37\times$ \\
\hline
\end{tabular}
\caption{Additional token cost multipliers over the Single-Agent
baseline on the four benchmarks not included in the main-text token
cost results.}
\label{tab:additional_token_cost}
\end{table*}

Table~\ref{tab:additional_token_cost} complements the token-cost
results reported in the main paper. The additional benchmarks exhibit
a similar overall pattern: interactive MAS methods require
substantially more tokens than Single-Agent, while higher token usage
does not necessarily produce stronger collaboration gains. MAPR
consistently incurs the largest cost among the evaluated methods.
CLEARS also requires a relatively large token budget, much of which is
introduced by cross-agent claim evaluation. Removing CCE reduces its
token cost by approximately half across the four benchmarks, while the
main results show that CCE consistently improves CG. These results
further support the main-text analysis of how additional computation
is converted into collaboration gains.

\end{document}